\documentclass[sigconf, nonacm]{acmart}

\usepackage{tikz}
\usepackage{amsmath}
\usepackage{float}
\usepackage{listings}
\usepackage{algorithm}
\usepackage[rightComments=true, indLines=true]{algpseudocodex}
\usepackage{graphicx}
\usepackage{textcomp}
\usepackage{xcolor}
\usepackage{subcaption}
\usepackage{caption}
\usepackage{circledsteps}
\usepackage{multirow}
\usepackage{booktabs}
\usepackage{tabularx}
\usepackage{balance}
\usepackage[subtle, lists=tight, mathdisplays=tight]{savetrees}
\usepackage{pdfpages}
\usepackage{circledsteps}
\usepackage[font=small,skip=3pt]{caption}

\newcommand{\scmacro}{{Quirinus}}
\newcommand{\olmacro}{{Minerva}}

\begin{document}
\title{Optimizing Geo-replicated Deterministic Databases}

%%
%% The "author" command and its associated commands are used to define the authors and their affiliations.
\author{Yunhao Mao}
\affiliation{%
  \institution{University of Toronto}
}
\email{yunhao.mao@mail.utoronto.ca}

\author{Shiquan Zhang}
\affiliation{%
    \institution{University of Toronto}
}
\email{shiquan.zhang@mail.utoronto.ca}

\author{Gengrui Zhang}
\affiliation{%
 \institution{Concordia University}
}
\email{gengrui.zhang@concordia.ca}

\author{Hans-Arno Jacobsen}
\affiliation{%
  \institution{University of Toronto}
}
\email{jacobsen@eecg.toronto.edu}

%%
%% The abstract is a short summary of the work to be presented in the

%% article.

\begin{abstract}

In transactional databases, achieving active-active replication among geo-distributed replicas is challenging due to the high coordination cost in ensuring the serializability of concurrent cross-node transactions. Deterministic concurrency control (DCC) has emerged as a promising alternative: it establishes a global order for a batch of transactions before execution and thereby amortizes this coordination cost across the batch. To better understand the performance characteristics of DCC, we conduct a theoretical analysis of traditional two-phase locking and optimistic concurrency control to account for geo-replication and introduce a model for DCC. We compare the three paradigms and identify the conditions under which DCC is advantageous. Guided by the insights from this analysis, we then present \scmacro{}, a DCC-based geo-replicated database that optimizes transaction execution through speculative execution, transaction reordering, integration of conflict-free replicated data types (CRDTs), and fast-path commit. Even without these optimizations, \scmacro{} achieves up to $6.2\times$ the TPC-C throughput of Tiga under our configuration.

\end{abstract}

\maketitle
\pagestyle{plain}

\section{Introduction}

Geo-replication is essential for globally scaled databases, as it provides extremely high availability to the applications that serve global users. Commonly, a geo-replicated OLTP database consists of transaction execution nodes and replicated storage nodes distributed across different data centers (DCs) for horizontal scalability and fault tolerance.

To provide \textit{serializable} isolation, databases use \textit{concurrency control} protocols such as two-phase locking (2PL) and optimistic concurrency control (OCC) to resolve conflicts and establish a serial transaction order. Under geo-replication, however, these protocols incur significant coordination overhead~\cite{zhang_tapir_2018} because a transaction cannot commit until its writes have been durably and consistently replicated. This guarantee is typically provided through \textit{consensus replication}~\cite{diegoRaft2014,lamportPaxosMadeSimple2001}, which adds cross-DC communication to the commit path~\cite{zhang2022bftsurvey}. Because round-trip times (RTTs) between geo-distributed DCs can reach hundreds of milliseconds~\cite{bailis2013hat}, this coordination lengthens transaction lifetimes. Both 2PL and OCC resolve conflicts at runtime for each transaction, so longer lifetimes increase exposure to conflicts~\cite{yu1993cc}: 2PL transactions hold locks longer, whereas OCC transactions are more likely to abort and retry.

The support for an \textit{active-active} architecture further complicates the design of concurrency control protocols because the storage layer must now accept both reads and writes at any replica, requiring \textit{multi-leader replication}~\cite{ActiveActiveGeodistributedRedis,zhougeogauss2023,maominerva2026} instead of routing all requests to a designated leader and only using replicas for backups. Active-active offers better redundancy in the case of leader failures or a network partition between execution nodes and storage nodes~\cite{mansouri2017data,zhang2022starry}. However, transactions may concurrently read and update different replicas, requiring execution nodes to coordinate to ensure serializability.

Many prior studies have proposed techniques to reduce the coordination overhead, such as consolidating atomic commitment and replication~\cite{mu_consolidating_2016, lusnow2016, gengTiga2025}, using OCC with a fast commit path~\cite{fanOceanVistaGossipbased2019, luEpochbasedCommitReplication2021, shenMakoSpeculativeDistributed2025,maominerva2026}, or relaxing consistency guarantees~\cite{liredblue2012, bailis2014coordination, gotsmanCauseImStrong2016, zhang_tapir_2018}. While some can reduce commit latency to as low as one RTT under ideal, conflict-free conditions, they cannot eliminate multi-node concurrency control coordination as it is not possible to determine whether a transaction conflicts with another without communication.

An alternative is to amortize the coordination cost. When multiple operations require the same fixed-cost step, such as I/O or network communication, executing that step once for a \emph{batch} amortizes the cost across the operations and improves throughput~\cite{helland1989batch}. Transaction processing systems already use batching for replication and validation. However, transaction ordering cannot be batched when each transaction is ordered independently at runtime.

\begin{table*}[t]
    \centering
    \caption{OLTP Systems Overview}
    \label{tab:cc-replication}
    \footnotesize
    \setlength{\tabcolsep}{4pt}
    \renewcommand{\arraystretch}{1.10}
    \begin{tabularx}{\textwidth}{@{}
        >{\raggedright\arraybackslash}p{0.09\textwidth}
        >{\raggedright\arraybackslash}p{0.155\textwidth}
        >{\raggedright\arraybackslash}p{0.135\textwidth}
        >{\raggedright\arraybackslash}p{0.115\textwidth}
        >{\raggedright\arraybackslash}p{0.145\textwidth}
        >{\raggedright\arraybackslash}X@{}}
        \toprule
        \textbf{System} &
        \textbf{CC protocol} &
        \textbf{Replication type} &
        \textbf{Partial vs.\ full} &
        \textbf{Ordering mechanism} &
        \textbf{Notes} \\
        \midrule
        Spanner~\cite{corbettSpannerGoogleGlobally2013} &
        2PL+MVCC &
        Leader--follower &
        Partial / configurable &
        Physical Timestamp &
        \\

        CockroachDB~\cite{taftCockroachDBResilientGeoDistributed2020a} &
        Timestamp Ordering + MVCC &
        Leader--follower &
        Partial / configurable &
        Physical Timestamp &
        Hybrid concurrency control using timestamps and optimistic execution \\

        TAPIR~\cite{zhang_tapir_2018} &
        OCC &
        Leaderless &
        Partial &
        Physical Timestamp &
        \\

        Mako~\cite{shenMakoSpeculativeDistributed2025} &
        OCC &
        Leader--follower &
        Partial &
        Local Time &
        Requires RDMA \\

        GeoGauss~\cite{zhougeogauss2023} &
        OCC &
        Multi-leader &
        Full &
        Local Time &
        \\

        Starry~\cite{zhang2022starry} &
        OCC &
        Multi-leader &
        Partial &
        Logical Clock &
        \\

        Tiga~\cite{gengTiga2025} &
        Timestamp Ordering + OCC &
        Leader--follower &
        Partial or full &
        Logical Clock &
        \\

        Minerva~\cite{maominerva2026} &
        OCC + Deterministic &
        Multi-leader &
        Full &
        Logical Clock &
        Deterministic only for aborted transactions \\

        Calvin~\cite{thomsonCalvinFastDistributed2012} &
        Deterministic &
        Leader--follower &
        Full &
        Local Time &
        \\

        SLOG~\cite{renslog2019} &
        Deterministic &
        Leader--follower &
        Full &
        Log Order &
        \\

        Detock~\cite{nguyen2023detock} &
        Deterministic &
        Leader--follower &
        Hybrid; optional full materialization &
        Dependency Graph &
        \\

        Aria~\cite{luAriaFastPractical2020} &
        Deterministic &
        CC-only design &
        N/A &
        Logical Clock &
        \\

        Harmony~\cite{laiHarmony2023} &
        Deterministic &
        Blockchain &
        Full &
        Logical Clock &
        Block-level ordering \\

        \textbf{\scmacro{}} &
        Deterministic &
        Multi-leader &
        Full &
        Logical Clock &
        \\
        \bottomrule
    \end{tabularx}

    \vspace{2pt}

    \begin{minipage}{\textwidth}
        \footnotesize\emph{Replication coverage.}
        \emph{Full} means that every region stores every database partition;
        \emph{partial} means that a shard or range may be placed only in
        selected regions. A full logical replica may still be partitioned
        across multiple nodes within a region.
    \end{minipage}
\end{table*}

Deterministic concurrency control (DCC) brings batching to transaction ordering. It predetermines a global order for a batch of transactions \textit{before} execution, thereby removing runtime order resolution while preserving serializability~\cite{thomsonCalvinFastDistributed2012, faleiroBOHM2015, faleiroPWM2017, thomsonddbv2010, abadiOverviewDeterministicDatabase2018, qinCaracal2021, luAriaFastPractical2020, hildredCaerusLowLatencyDistributed2023, wang_gria_2023, hanblockchaindcc2025, hongHDCC2025}. It thus removes the need for strongly consistent replication at the storage layer; instead, the replication moves to the \textit{execution layer} that replicates transaction \textit{inputs} and runs the same deterministic execution at each replica. By decoupling ordering from execution, DCC amortizes the ordering coordination cost across a batch instead of paying it for each transaction. However, this design introduces two costs: (1) every transaction incurs batching delay, even without contention; and (2) execution is repeated at each replica, making execution efficiency critical.

Table~\ref{tab:cc-replication} summarizes the concurrency control and replication paradigms of representative OLTP systems with replication in mind. We observe that OCC has become a popular choice for geo-replication, while DCC is gaining traction in recent years. Although the strengths and weaknesses of these concurrency control paradigms are intuitively clear, we need to understand their advantageous regimes and trade-offs. Thus, we conduct a theoretical analysis that models and compares the commit latency and throughput of 2PL, OCC, and DCC. DCC is found to achieve favorable performance when the network RTT dominates execution time, whereas 2PL and OCC can suffer from unbounded latency under high contention. The replication speed and deterministic execution time are still the bottleneck for DCC, which requires optimization.

Guided by this analysis, we select four techniques to optimize DCC: speculative execution, transaction reordering, conflict-free replicated data types (CRDTs)~\cite{shapiro2011crdt,nuno2018newcrdt}, and fast-path commit. We implement these techniques in \scmacro{}, a DCC-based database designed for active-active geo-replication.\footnote{Source code: \url{https://github.com/yunhaom94/Quirinus-DCC}.}

The core idea that allows these optimizations to be realized is to synchronize replica state \textit{only} at batch boundaries, creating a \textit{consistent snapshot}~\cite{eldeeb_chardonnay_2023} between consecutive batches. This design allows transactions within a batch to be freely reordered to reduce the total execution cost, provided that all replicas derive the same deterministic order. For example, a transaction with a read--write dependency can instead read from the snapshot, allowing it to execute concurrently with the writer.

In summary, the contributions are as follows:
\begin{enumerate}

    \item We present explanatory models of 2PL, OCC, and DCC response time and saturation throughput and use them to identify qualitative regimes of contention and network delay in which each paradigm is advantageous.
    
    \item We present \scmacro{}, a database that utilizes DCC with optimizations that reduce the execution delay of DCC.
    
    \item We evaluate the three paradigms with simulated geo-distributed deployments to test the model's bottleneck predictions. Even without these optimizations, \scmacro{} achieves up to $6.2\times$ the TPC-C throughput of Tiga~\cite{gengTiga2025} under our configuration.
\end{enumerate}

The rest of this paper is organized as follows. Section~\ref{sec:bg} introduces the main paradigms of distributed concurrency control. Section~\ref{sec:perf} models the commit latency and throughput of these paradigms under geo-replication and compares their behavior under contention. Section~\ref{sec:cl} presents the design of \scmacro{}. Section~\ref{sec:eval} evaluates \scmacro{} on geo-distributed workloads. Section~\ref{sec:related} discusses related work.

\section{Distributed Concurrency Control}
\label{sec:bg}
Concurrency control protocols are well-studied in database systems~\cite{thomasian1998cc,yu1993cc,abadiOverviewDeterministicDatabase2018}. However, distribution and replication significantly complicate their design and implementation. We discuss three common paradigms of concurrency control in the context of active-active geo-replication.

\textbf{\textit{Pessimistic Concurrency Control (PCC).}}
PCC utilizes locks, such as the classical two-phase locking (2PL), to control access to shared data~\cite{ozsuPrinciplesDistributedDatabase2020, harding2017evaluation}. A transaction must acquire all necessary locks before it can commit (otherwise, any read or write operation will block on the lock acquisition), and it holds these locks until the end of the transaction. In 2PL, read operations acquire shared locks and write operations acquire exclusive locks to allow concurrent reads but serialized writes. The lock is usually implemented at the replicated storage layer if leader-based replication is used, because if transactions only attempt to acquire locks on the single leader replica, the ownership can be easily resolved locally by the leader.

In an active-active setting, the difficulty lies in establishing a global lock order among the execution nodes through some form of centralized or distributed lock authority: a replicated lock manager, or a quorum/consensus layer that makes lock ownership globally visible~\cite{bernstein1981cc, gray1996replication}, adding another consensus round in addition to the consensus used for data replication. As a result, 2PL is the most sensitive of the three paradigms to network delay across replicas, since every lock acquisition requires cross-DC communication.

\textbf{\textit{Optimistic Concurrency Control (OCC).}}
OCC splits a transaction into read, validation, and write phases. Execution is never blocked, reads are served from the locally visible copies of the data, and the transaction may expose its writes only after validation succeeds~\cite{ozsuPrinciplesDistributedDatabase2020, harding2017evaluation}. Validation checks whether the execution can be serialized with previously validated transactions, and invalid transactions are aborted and retried. The validation is typically conducted via a timestamp protocol to determine whether another transaction makes an update to the values that the transaction is accessing.

OCC fits naturally with geo-distributed replication, because execution can proceed locally at one replica and global coordination can be deferred until commit so that the coordination cost is paid per transaction instead of per lock~\cite{mahmoud2014maat}. Compared to 2PL, OCC has become the more common paradigm for geo-distribution as its validation phase also offers more room for optimization, such as reducing the cost of 2PC as in TAPIR~\cite{zhang_tapir_2018}, epoch-based validation~\cite{luEpochbasedCommitReplication2021}, and speculative commit~\cite{shenMakoSpeculativeDistributed2025}.

Aborts and retries, however, remain expensive, as each one imposes additional cost. Because the timestamp must be globally consistent, replication prolongs the time that a transaction is exposed to conflict, thus increasing the abort probability in geo-replication.

\textbf{\textit{Deterministic Concurrency Control (DCC).}}
\label{sec:bg:dcc}
DCC avoids the runtime order resolution required by PCC and OCC. Instead of letting transactions execute concurrently and resolving conflicts at runtime, DCC builds a \textit{dependency graph statically} before execution and then executes the transactions in this predetermined order~\cite{abadiOverviewDeterministicDatabase2018}. 

\begin{figure}[h]
    \centerline{\includegraphics[scale=1] {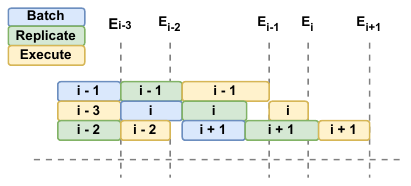}}
    \caption{Pipeline of a DCC protocol.}
    \label{fig:pipeline}
\end{figure}

When used with geo-replication, a database that uses a DCC protocol no longer requires the storage layer to handle replication; instead, the execution layer replicates the transaction inputs such as SQL queries or parameters to execution nodes in other DCs. The database can essentially be viewed as a replicated state machine, and states are maintained by local storage~\cite{pedone2003database}. An active-active architecture can be easily achieved as long as the replication protocol is a multi-leader or leaderless one~\cite{kleppmannDesigningDataIntensiveApplications2016} that accepts requests at any replica. A DCC protocol such as Calvin~\cite{thomsonCalvinFastDistributed2012} generally follows a three-stage pipeline, as shown in Figure~\ref{fig:pipeline}:

\begin{enumerate}

    \item \textbf{Batching:} Transactions are collected at each replica and grouped into batches. Each transaction is assigned a globally unique sequence number. The batch size is a parameter tuned for pipeline utilization, and each batch constitutes an \emph{epoch}.

    \item \textbf{Replication:} Once a batch has been collected, it is replicated to all replicas using a consensus protocol.

    \item \textbf{Deterministic Execution:} Each replica executes the transactions in the order given by their pre-established sequence numbers.
\end{enumerate}

Figure~\ref{fig:pipeline} also illustrates both the case where replication takes longer than execution and the case where execution takes longer than replication. Epoch $i-1$ is the first case where replication is the bottleneck (slower than the execution of epoch $i-2$), so the execution of epoch $i-1$ must wait for its replication to finish. The execution of epoch $i-1$ is, in turn, longer than the replication of epoch $i$, so the execution of epoch $i$ must wait for the execution of epoch $i-1$ to finish.

The replication step is bounded by the network RTT, which is the physical limit of the system. The batch size can be tuned based on the duration of the other stages. The execution step is therefore a primary target for optimization, as it must be fast enough to reach optimal throughput, while the replication step must also be optimized to increase the overall performance ceiling of the system. Because contention no longer grows asymptotically with network latency, the coordination delay can be amortized across a batch of transactions, which is advantageous under geo-replication.

The downside of pre-ordering and batching is that the preprocessing delay must always be paid, even if there is no contention, which adds unavoidable latency to every transaction and may prolong the response time of short transactions that have few conflicts. Another common issue is the lack of support for interactive transactions, since the read and write keys must be declared in advance for sequencing~\cite{abadiOverviewDeterministicDatabase2018}.

\section{Modeling Latency and Throughput}
\label{sec:perf}

To better quantify the performance characteristics of these parad\-igms, we present an analytical model of the latency and throughput of the different concurrency control protocols under active-active geo-replication. We assume full database replication (no cross-DC reads), fault tolerance through consensus, and fixed-length transactions in which each transaction accesses the same number of records.

\paragraph{Preliminaries and Notation}
We adapt the classic analytical model of transaction processing by Yu et al.~\cite{yu1993cc} for the expected transaction response time (end-to-end latency). We then extend this model in three ways: adding coordination cost via cross-datacenter network delay ($\tau$), modeling DCC response time and throughput, and deriving the theoretical saturation throughput (maximum stable throughput) of all CC paradigms. Table~\ref{tab:notation} summarizes the notation used in the rest of the paper.

Yu et al. give the response time of a transaction for 2PL: $$R_{2PL} = R_{\text{INPL}} + R_E + N_L P_W R_W + T_{\text{Commit}},$$ and for OCC: $$R_{OCC} = R_{\text{INPL}} + R_E + \frac{P_A}{1-P'_A}\,(T_{\text{Backoff}} + R'_E) + T_{\text{Commit}}.$$

We simplify some parameters to focus on the impact of the coordination delay. The CPU and I/O times are combined into a single variable $D = R_{\text{INPL}} + N_L R_0$, where $R_0$ is the average per-granule access time (i.e., $a$ in 2PL and $\hat R$ in OCC). Following Yu's model, the compute resource is treated as an open queue with arrival rate $\lambda$, served by $K$ CPUs, where $D$ is the time to process one transaction on a CPU. The compute utilization is therefore $\rho = \lambda D / K$ (ignoring locks). Under the fixed-length transaction assumption, each transaction sees $\lambda N_L^2 T_H / L$ expected conflicts under both 2PL and OCC, given $\lambda T_H$ concurrent in-flight transactions (by Little's Law) and a per-pair conflict probability of $p_c \approx N_L^2/L$.

For 2PL and OCC, the write set must be replicated before commit, and we assume for simplicity that the local commit time is negligible compared to $\tau$, so that $T_{\text{Commit}}
= T_{\text{Commit}}^{\text{local}} + \tau
\approx \tau$. For DCC, all replicas execute the same replicated transactions, so no coordination time is paid at commit.

\begin{table}[ht!]
\centering
\caption{Notation used in the analytical model. $\star$ denotes ``knob'' parameters that we focus on in the analysis.}
\label{tab:notation}
\small
\begin{tabular}{@{}c@{\hspace{0.6em}}lp{0.66\linewidth}@{}}
\toprule
& \textbf{Symbol} & \textbf{Meaning} \\
\midrule
\multicolumn{3}{@{}l}{\textbf{Adopted from Yu et al.}} \\
\multirow{10}{*}{\rotatebox[origin=c]{90}{\textbf{Common}}}
& $L$ & Number of granules (data objects) in the database \\
& $N_L$ & Granules accessed per transaction (fixed length) \\
& $\boldsymbol{\lambda}$ & \textbf{Transaction arrival rate$\boldsymbol{\star}$}  \\
& $\boldsymbol{K}$ & \textbf{Number of CPUs$\boldsymbol{\star}$} \\
& $R$ & Mean transaction response time \\
& $R_{\text{INPL}}$ & Mean setup time per transaction \\
& $R_E$ & Mean execution time (compute considered) \\
& $T_{\text{Commit}}$ & Mean commit time including results replication \\
& $T_{\text{Backoff}}$ & Mean backoff time after abort \\
& $T_H$ & Mean granule holding time \\
\midrule
\multirow{6}{*}{\rotatebox[origin=c]{90}{\textbf{2PL}}}
& $a$ & Mean lock-holding time in the execution substate   \\
& $b$ & Unconditional lock-acquisition time, $b = P_W R_W$ \\
& $P_W$ & Per-lock-request contention probability \\
& $P_W^T$ & Per-transaction contention, $P_W^T = P_W(N_L-1)$ \\
& $R_W$ & Mean lock-waiting time given contention \\
& $G$ & Sum of mean lock-holding times on all granules \\
\midrule
\multirow{3}{*}{\rotatebox[origin=c]{90}{\textbf{OCC}}}
& $P_A, P'_A$ & Per-transaction abort probability (first \& re-run) \\
& $\hat R, R'$ & Mean per-state execution time (first \& re-run) \\
& $R'_E$ & Mean re-execution time on rerun, $R'_E = N_L R'$ \\
\midrule
\multicolumn{3}{@{}l}{\textbf{Introduced in our work}} \\
& $\boldsymbol{\tau}$ & \textbf{Mean delay of RTT-scaled coordination among replicas, $\boldsymbol{O(RTT)}\boldsymbol{\star}$} \\
& $\boldsymbol{D}$ & \textbf{Transaction compute demand, without considering compute resources$\boldsymbol{\star}$} \\
& $\boldsymbol{p_c}$ & \textbf{Pairwise conflict probability under uniform access, $\boldsymbol{p_c \approx N_L^2/L}\boldsymbol{\star}$} \\
& $\boldsymbol{\lambda^*}$ & \textbf{Saturated (maximum stable) throughput$\boldsymbol{\star}$} \\
\bottomrule
\end{tabular} 
\end{table}

\subsection{Pessimistic Concurrency Control (2PL)}
\paragraph{Modeling Latency}

For locking-based PCC, lock-wait time is the main source of latency. A ``blow-up'' of lock wait may occur when transactions wait for a lock that is itself blocked by another transaction, producing a cascading effect. Yu et al. show that if the average number of lock contentions per transaction exceeds $P_W^T \approx 0.75$, the lock wait grows unbounded. This bound is obtained by equating two expressions for $R_W$: one from the definition $b = P_W R_W$, and one from the construction of $R_W$ for a single lock wait. The result is quadratic in $b$, because $R_W$ depends on itself, and a real, positive solution (a finite lock wait) exists only when $P_W^T \le 0.75$.

Under geo-replication, every lock request needs at least one RTT to confirm whether the lock is granted~\cite{nawabhelios2015}, so the \textit{mean unconditional lock-acquisition time} for each lock is approximately one RTT in the absence of contention. We therefore define $b' = \tau + P_W R_W$. This extra delay lengthens how long each transaction holds its locks, which in turn raises the chance that other transactions collide ($P_W$). Adding $\tau$, we obtain the response time formula:

\[
\begin{aligned}
R_{\text{2PL}}
&= R_{\text{INPL}} + N_L\,a + N_L b' + T_{\text{Commit}} \\
&\approx R_{\text{INPL}} + N_L\,a + N_L(\tau + P_W R_W) + \tau \\
&= R_{\text{INPL}} + N_L\,a + N_L P_W R_W + (N_L+1)\tau.
\end{aligned}
\]

Following Yu et al.'s lock wait threshold derivation, we estimate the threshold after adding $\tau$:

$$
\;P_{W,\text{MAX}}^T \;=\; \frac{0.75}{\,1 + \tau/a\,}.
$$

This means that the blow-up threshold is lowered as $\tau$ increases. See Appendix~\ref{app:blowup} for the detailed derivation.

\paragraph{Modeling Throughput}
The maximum stable throughput in 2PL is determined by both the blow-up threshold and the compute resource
$
\lambda^*_{\text{2PL}}
\;=\; \min\!\bigl(\lambda^*_{\text{CPU}},\, \lambda^*_{\text{Cont}}\bigr)
$.

The compute-bound throughput is simply $\lambda^*_{\text{CPU}} \;=\; \frac{K}{D}$ under an M/M/K queueing model in the absence of contention.

For the lock-wait blow-up bound, we first get $b'_{\text{Cont}} \;=\; a + 2\tau$ when $P_W^T = 0.75/(1+\tau/a)$. Then, we substitute $b'_{\text{Cont}}$ into the formula for $G'$ ($G$ with $b'$) to get $G'_{\text{Cont}}$~\footnote{In Yu et al.'s model, each transaction obtains locks on each granule in a serialized manner and proceeds through a linear sequence of states. Each acquired lock transitions the state of a transaction from $\tilde i$ to $\hat i$, where $a$ is the mean lock-holding time. $G$ is the total lock-holding time of a transaction. Appendix~\ref{app:lock-exposure} gives the full derivation.}:

$$
G'_{\text{Cont}}
\;=\; \tfrac{N_L(N_L{+}1)}{2}\,a + \tfrac{N_L(N_L{-}1)}{2}\,(a+2\tau) + N_L \tau \;=\; N_L^2(a+\tau).
$$

Using the definition of $P_W^T$ and multiplying by $(N_L{-}1)$ gives the mean number of contentions per
transaction:

$$
P_W^T
\;=\; P_W(N_L{-}1)
\;=\; \lambda (N_L{-}1)\frac{G'}{L}.
$$

Replacing $G'$ with $G'_{\text{Cont}}$ and solving for $\lambda^*_{\text{Cont}}$:

$$
\lambda^*_{\text{Cont}}(N_L{-}1)\frac{G'_{\text{Cont}}}{L}
\;=\; \frac{0.75}{1+\tau/a}.
$$

$$
\lambda^*_{\text{Cont}}
\;=\; \frac{0.75\,L\,a}
           {N_L^2\,(N_L{-}1)\,(a+\tau)^2}.
$$

\subsection{Optimistic Concurrency Control}
We assume that initial attempts and reruns have the same abort probability and compute demand, so $P'_A=P_A$ and both use $D$~\footnote{Yu et al. model buffering for reruns, which results in a different computation cost; we assume that is negligible in a geo-distributed setting.}.

\paragraph{Modeling Latency}
For OCC, the main source of overhead is aborts and retries. Although OCC does not block, a transaction is \emph{exposed} to conflicts from its first read until the end of commit, and any conflicting commit during this window causes the transaction to abort. We assume that validation and write-set replication happen in the same $\tau$ round. Following the realistic pure-OCC approximation of Yu et al., the coordination delay extends the commit window and thus increases the abort probability for both the first run and any rerun:

$$
P_A \;\approx\; 1 - \exp\!\bigl(-\lambda N_L^2 T_H/L\bigr)
       \left(1 - \frac{\lambda N_L \tau}{L}\right)^{N_L}.
$$

Its $O(1/L)$ linearization is:
$$
P_A \;\approx\; \frac{\lambda N_L^2(T_H+\tau)}{L}.
$$

The overall response time combines the coordination round for the first attempt and each retry:
$$
R_{OCC} \;=\; R_{\text{INPL}} + N_L \hat R + \tau
    \;+\; \tfrac{P_A}{1-P'_A}(T_{\text{Backoff}} + N_L R' + \tau).
$$

Note that OCC has no blow-up threshold: the abort probability does not depend on itself, only on the number of concurrent transactions in the system, and adding $\tau$ does not change this linearity.
\paragraph{Modeling Throughput}
Since OCC is bounded only by compute resources, the throughput is determined by the number of aborts and retries. Thus, we calculate the expected number of execution attempts per transaction. With the $O(1/L)$ linearized approximation for the abort probability, we define $P_A \approx \lambda x$, where $x = N_L^2(T_H+\tau)/L$ is the abort coefficient after adding validation and replication coordination delay. The expected number of retries is $\frac{\lambda x}{1-\lambda x}$; including the first attempt, the expected compute demand is $\frac{D}{1-\lambda x}$.

We then apply the M/M/$K$ utilization condition
$\rho = \frac{\lambda D}{K(1-\lambda x)} < 1$.
Substituting $x$ gives the maximum stable throughput:
$$
\lambda^*_{\text{OCC}}
   = \frac{K}{D+Kx}
   = \frac{K}{D + K\,N_L^2(T_H+\tau)/L}
   = \frac{KL}{LD + K\,N_L^2(T_H+\tau)}.
$$

\subsection{Deterministic Concurrency Control}
\label{sec:dcc}
\paragraph{Modeling Latency}
\label{sec:dcc-latency}
We extend the framework of Yu et al. to DCC and model a system that uses a Calvin-like distributed concurrency control protocol~\cite{thomsonCalvinFastDistributed2012}. As discussed in Section~\ref{sec:bg:dcc}, DCC consists of a three-stage pipeline: batching, replication, and execution. We combine the batching and replication stages into a single ``sequencing'' queue that pays $\tau$ for coordination, since the batch interval is a tunable parameter that directly affects the replication throughput, as we discuss shortly. Execution is a separate queue that processes replicated transactions with $K$ parallel threads and a service time of $D$. DCC performs no post-execution write-set replication. Finally, we assume that all replicas have identical transaction execution times.

We define the commit response time as:
$$
R \;=\; \underbrace{\Delta_B/2}_{\text{batch wait}} \;+\; \underbrace{\tau}_{\text{replication time}} \;+\; \underbrace{W_E}_{\text{exec wait}} \;+\; \underbrace{D}_{\text{own exec}}.
$$

Here, $\Delta_B$ is the fixed time interval used to seal a batch. DCC can encounter a consensus delay blow-up when $\Delta_B < \tau$, because batches then arrive at the consensus layer faster than it can replicate them, causing the sequencing queue to grow without bound. We therefore assume $\Delta_B \geq \tau$ for a stable system, with transactions still entering the execution queue at rate $\lambda$. Note that we do not use $R_E$ to denote the execution time, as we did for 2PL and OCC, because any computational wait time is absorbed into $W_E$, leaving $D$ as the unconditional compute time required for a transaction.

Next, we derive $W_E$, the mean waiting time when a transaction $T$ arrives in the stable state with $N_E$ transactions in the execution queue. Because of determinism, $T$ must wait for all previous transactions that conflict with $T$, as well as their predecessors, to finish execution. We need to consider the longest dependency chain of $T$, because all transactions in the chain are executed serially.

We model this by constructing a random directed acyclic graph (DAG) over the partially ordered set of transactions to represent the dependencies among transactions. Assume there are $N_E$ vertices, where each vertex is a transaction labeled with a $tid$. Each pair of vertices is connected independently with the same probability $p_c$, and the edge always points from the higher $tid$ to the lower $tid$. When $T$, which has the highest $tid$, is inserted into the DAG, it again connects to each existing vertex with probability $p_c$.

Define $I$ as the length of the longest dependency chain of $T$ and $S$ as the set that $T$ depends on; then we get:

$$
I \;=\;
\begin{cases}
0, & S = \emptyset, \\
1 + \max_{i \in S} I_{T_i}, & S \neq \emptyset
\end{cases}.
$$

Defining $\ell(i - 1) = E\bigl[I_i\bigr]$ and $M \equiv \max\{i : T \to T_i\}$, we apply a \emph{mean-field approximation}~\cite{wainwrightGraphicalModels2008} to obtain a closed-form expression. We replace each random $I_{T_i}$ inside the maximum by its mean $\ell(i{-}1)$ to obtain a closed-form length. This approximation treats the longest path as the longest of the mean path lengths and assumes that the highest-index child is the most likely to lie on the longest path:

$$
E\!\bigl[\,\max_{i:\,T \to T_i} I_{T_i}\,\bigr]
\;\approx\;
\max_{i:\,T \to T_i} \ell(i - 1)
\;=\; \ell(M - 1).
$$

Then, for any $N$ transactions and any transaction $T_m$ with $m \le N$, we have:
$$
\Pr(M = m) \;=\; p_c\,(1 - p_c)^{N - m},
\qquad \Pr(M = 0) \;=\; (1 - p_c)^N;
$$

$$
\ell(N) \;=\;
\sum_{m=1}^{N} p_c\,(1 - p_c)^{N - m}\,\bigl(1 + \ell(m - 1)\bigr).
$$

Solving this mean-field recurrence gives (see Appendix~\ref{app:longestpath} for the detailed derivation):

$$
\ell(N) \;=\; p_c\,N
\qquad \ell(N_E) \;=\; N_E\,p_c.
$$

Finally, the wait time is the maximum of the execution queue wait (the computation bound) and the dependency chain wait (the contention bound):

$$
W_E \;=\; \max\!\bigl(\underbrace{\tfrac{N_E\,D}{K}}_{\text{queue wait}},\; \underbrace{p_c\,N_E\,D}_{\text{chain wait}}\bigr)
      \;=\; N_E\,D \cdot \max\!\bigl(1/K,\, p_c\bigr).
$$

In the stable state, $N_E$ is in turn determined by Little's Law as $N_E \;=\; \lambda \,(W_E + D)$.

\paragraph{Modeling Throughput}
Since DCC has neither reruns nor lock waits, its throughput is bounded only by the speeds of the sequencing queue and the execution queue. As noted above, if the batch interval is shorter than the replication delay, the sequencing throughput is $\lambda_{\text{seq}} = \frac{\lambda \Delta_B}{\tau}$; otherwise, the sequencing queue is not a bottleneck and $\lambda_{\text{seq}} = \lambda$.

For the execution queue, we can reverse-engineer the stable $N_E$. First, define $ K_{\text{eff}} \equiv \min\!\left(K,\,\frac{1}{p_c}\right)$, then:

$$N_E = \lambda D\!\left(\frac{N_E}{K_{\text{eff}}}+1\right);  \qquad
N_E^*
  =
  \frac{\lambda D}
       {1-\lambda D/K_{\text{eff}}}.$$

A finite, positive pool requires a positive denominator, that is, $\lambda D/K_{\text{eff}}<1$. The saturated rate of the execution stage is therefore:

$$
\lambda^*_{\text{exec}}
  = \frac{K_{\text{eff}}}{D}
  = \frac{\min(K,\,1/p_c)}{D}
  = \min\!\left(\frac{K}{D},\,\frac{1}{p_cD}\right).
$$

Finally, the overall saturated throughput is the minimum over the sequencing and execution stages:
$$
\lambda^*_{\text{DCC}}
  = \min\!\left(\lambda_{\text{seq}},\, \lambda^*_{\text{exec}}\right)
  = \min\!\left(\frac{\lambda \Delta_B}{\tau},\, \frac{\min(K,\,1/p_c)}{D}\right).
$$

\subsection{Comparisons}
\label{sec:comparison}
In this section, we compare the advantageous regimes for coordination delays and contention levels among the three paradigms with respect to saturated throughput. We treat the other parameters as constants and focus on $\tau$ and $p_c$ only.

To compare throughput, we standardize the contention parameter to $p_c$, as in Section~\ref{sec:dcc}, and simplify the compute demand to $D$ for all paradigms (ignoring the slowdown in the execution queue beyond full utilization, since we are only interested in the throughput ceiling). This gives $a \approx D/N_L$, and for OCC $\hat R \approx D/N_L$, $T_H = (N_L{+}1)\hat R/2 \approx D/2$, and $x = N_L^2(T_H{+}\tau)/L = p_c\,(T_H + \tau)$. Substituting these into the throughput expressions gives:

\begin{figure*}[!t]
    \centering
    \begin{subfigure}[t]{0.32\textwidth}
        \includegraphics[width=\textwidth]{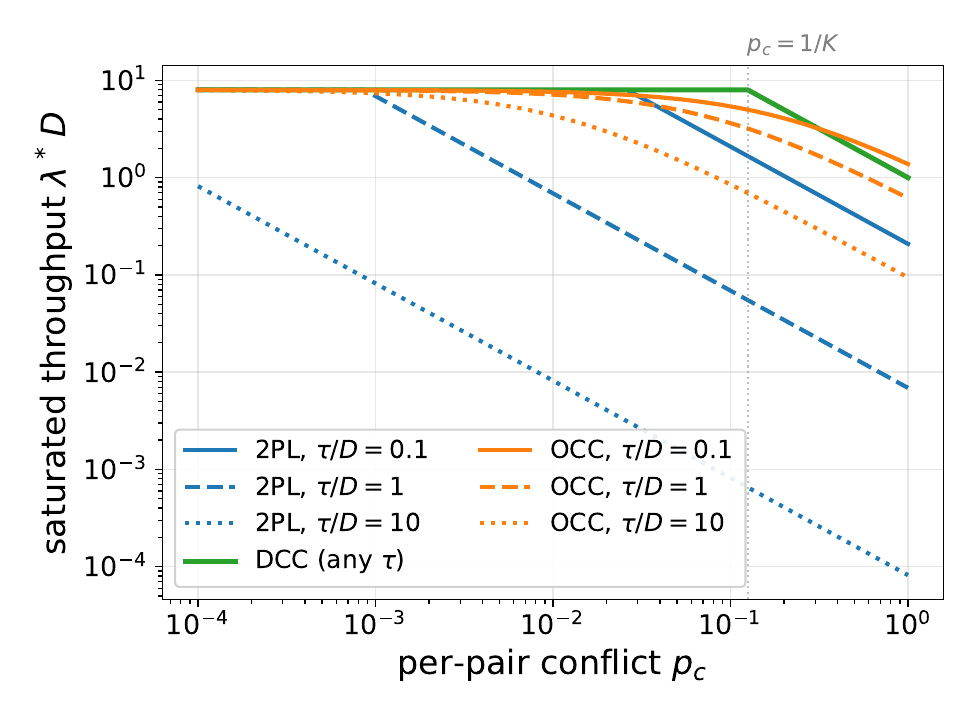}
        \caption{Saturated throughput $\lambda^* D$ versus per-pair conflict probability $p_c$. In DCC, $p_c=1/K$ marks the shift from the compute bound to the dependency-chain bound.}
        \label{fig:cc-throughput}
    \end{subfigure}\hfill
    \begin{subfigure}[t]{0.32\textwidth}
        \includegraphics[width=\textwidth]{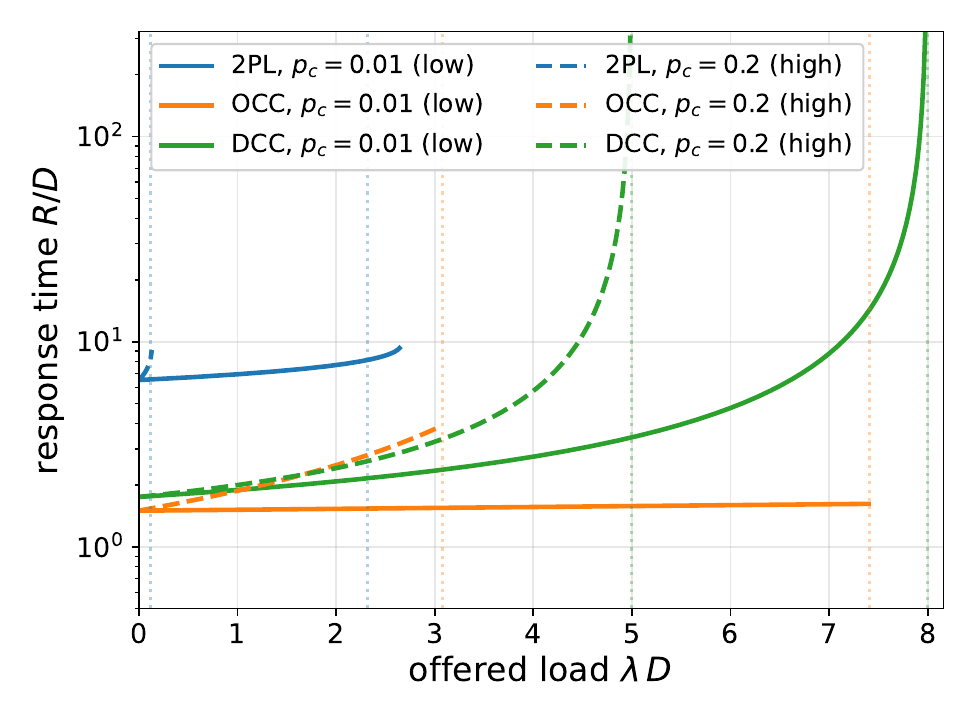}
        \caption{Response time $R/D$ vs.\ input throughput at $\tau/D = 0.5$ (low latency).}
        \label{fig:cc-resp-small}
    \end{subfigure}\hfill
    \begin{subfigure}[t]{0.32\textwidth}
        \includegraphics[width=\textwidth]{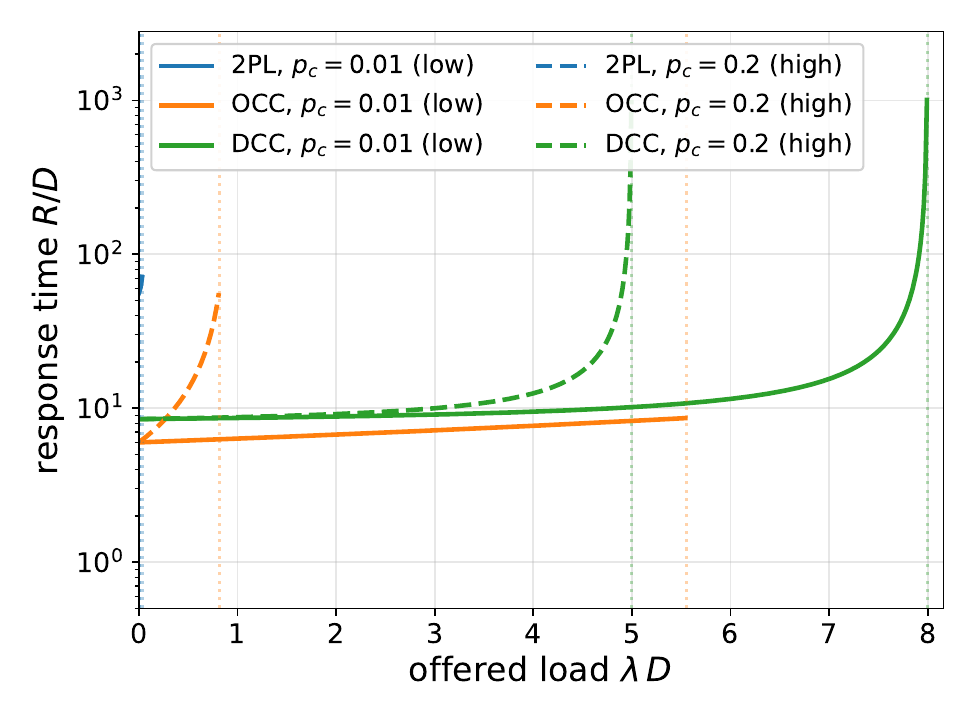}
        \caption{Response time $R/D$ vs.\ input throughput at $\tau/D = 5$ (high latency).}
        \label{fig:cc-resp-large}
    \end{subfigure}
    \vspace{0.4cm}
    \caption{Analytical comparisons of CC paradigms with $K=8$, $N_L=10$, and negligible local commit time relative to cross-DC delay. We nondimensionalize $\tau$ and $\lambda$ by $D$ to show the relative impact of coordination delay and contention. For DCC, we show only the execution-stage bound and set the batch interval to $\Delta_B = \tau$, the smallest value satisfying the stability requirement $\Delta_B \geq \tau$.}
    \label{fig:cc-comparison}
\end{figure*}

\paragraph{For 2PL}
$$
\lambda^*_{\text{lock}}
   \;=\; \frac{0.75\,N_L\,D}{p_c\,(N_L{-}1)\,(D + N_L\tau)^2}.
$$

For large $N_L$, using $(N_L{-}1) \approx N_L$ in the leading factor gives:

$$
\lambda^*_{\text{lock}}
   \;\approx\; \frac{0.75\,D}{p_c\,(D + N_L\tau)^2}.
$$

\paragraph{For OCC}

$$
\lambda^*_{\text{OCC}}(\tau,p_c)
   \;=\; \frac{K}{D + K\,p_c\,(D/2 + \tau)}.
$$

Next, we compare each pair of the paradigms.

\paragraph{OCC vs. DCC}

When DCC is compute-bound:

$$
\frac{K}{D + K p_c(D/2 + \tau)} \;=\; \frac{K}{D},
$$

whose only solutions are $p_c = 0$ and $\tau = -D/2$. For contention-bound DCC,

$$
\frac{K}{D + K p_c(D/2 + \tau)} \;=\; \frac{1}{p_c D}.
$$

Solving for $\tau$ gives:

$$
\tau^{\text{O/D}}
   \;=\; D\!\left(1 - \frac{1}{K p_c}\right) - \frac{D}{2}
   \;=\; D\!\left(\tfrac{1}{2} - \tfrac{1}{K p_c}\right).
$$

For contention-bound DCC, OCC achieves higher saturated throughput only when $\tau < \tau^{\text{O/D}}$; the two tie at equality. For large $Kp_c$, this threshold approaches $D/2$ from below. Thus, OCC wins only when $\tau < D/2$, i.e., when the coordination delay is shorter than half the transaction's compute time. When contention is low ($0 < p_c \leq 1/K$), DCC achieves higher saturated throughput because it avoids retry work; their throughput limits coincide at $p_c=0$. These comparisons assume that sequencing does not limit DCC throughput.

\paragraph{2PL vs. DCC}

When DCC is compute-bound:

$$
\frac{0.75\,D}{p_c\,(D + N_L\tau)^2} \;=\; \frac{K}{D}.
$$

Taking the positive root for $\tau$ gives:
$$
\tau^{\text{2/D}}
   \;=\; \frac{D}{N_L}\!\left(\sqrt{\tfrac{0.75}{K p_c}} - 1\right).
$$

When $K p_c < 0.75$, 2PL ties with DCC if $\tau < \tau^{\text{2/D}}$. Otherwise, 2PL can reach the lock wait blow-up threshold even at $\tau = 0$, so DCC always wins.

When DCC is contention-bound:
$$
\frac{0.75\,D}{p_c\,(D + N_L\tau)^2} \;=\; \frac{1}{p_c D};
$$
$$
D + N_L\tau \;=\; \sqrt{0.75}\,D.
$$
Since $p_c$ cancels out and $\sqrt{0.75} < 1$, we cannot find a non-negative $\tau$ that satisfies the equation. This means that 2PL is again always lower than DCC in the contention-bound regime, and the gap grows as $\tau$ increases.

\paragraph{2PL vs. OCC}
Let:
$$
\frac{0.75\,D}{p_c (D + N_L\tau)^2} \;=\; \frac{K}{D + K p_c(D/2 + \tau)}.
$$

Setting $\tau = 0$ reduces this equation to the traditional comparison between 2PL and OCC: 2PL performs better when contention is low, while OCC performs better when contention is high, matching the results of Yu et al.

With a large $\tau$, the $1/\tau$ vs. $1/\tau^2$ scaling means that OCC eventually outperforms 2PL for any fixed $p_c>0$. The product $Kp_c$ is the useful regime parameter because it approximates the number of direct conflicts a transaction sees when the system tries to run $K$ transactions in parallel.

\paragraph{Summary}

Figure~\ref{fig:cc-comparison} shows the theoretical saturated throughput and response times of the three paradigms. 2PL is sensitive to coordination delay because each lock acquisition requires cross-replica communication. OCC offers the lowest response time at low load, when queueing and retries are negligible, because it avoids DCC's batching delay. When sequencing does not limit throughput, DCC achieves higher saturated throughput than OCC under low but nonzero contention. DCC also outperforms both OCC and 2PL when contention and coordination delay are high.

An important caveat is that, although we assume $\tau$ is the same for all paradigms, in practice $\tau$ is not a constant cost but depends heavily on the coordination or consensus protocol. This motivates the design of \scmacro{}, which optimizes replication and execution to minimize both the sequencing time and the execution delay.

\section{\scmacro{} Design}
\label{sec:cl}
To explore optimization opportunities for DCC based on the insights, we present \scmacro{}, a deterministic database that targets \textbf{active-active, geo-replicated} deployments. We highlight the following design features:

\begin{itemize}
    \item \textbf{Lightweight Synchronization}: Replica state is materialized only at \emph{consistent snapshots} rather than at individual transactions, using a single synchronization barrier at the batch level. At the serializable isolation level, this guarantees serializable execution and identical snapshots across replicas for each completed batch. Synchronous coordination occurs only once on the critical path of a transaction, so that the coordination latency is \emph{amortized across the batch}.

    \item \textbf{Speculative Execution}: We show that the throughput of DCC can be bounded by compute power for both dependency-chain and non-conflicting transactions. By allowing transactions to execute speculatively against the snapshot before commit, we reduce the effective execution time $D$, which \emph{improves throughput under low contention and reduces commit latency}.
    
    \item \textbf{Conflict Avoidance}: Because the dependency chain is another main constraint in DCC, we adopt conflict-avoidance techniques, such as adjustable isolation levels, commutative operators with CRDTs, and transaction reordering, to further \emph{reduce contention overhead}.
    
    \item \textbf{Fast-path Commit}: To further \emph{reduce commit latency} for non-conflicting transactions, we allow some transactions, such as read-only transactions, to respond as successful without waiting for the synchronization barrier, provided that they are guaranteed to be valid.
\end{itemize}

\subsection{Design Overview}

\begin{figure}[h]
    \centerline{\includegraphics[scale=0.70] {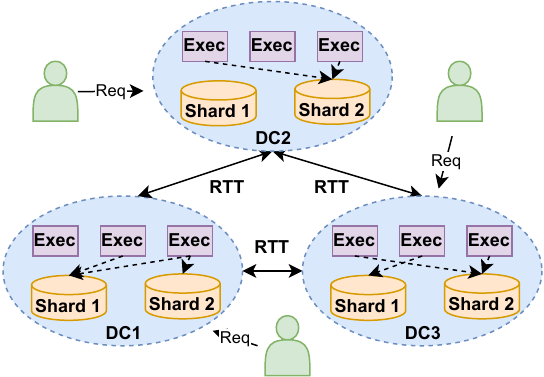}}
    \caption{System architecture. Each DC is a replica with execution nodes and partitioned storage; clients submit transactions to any DC.}
    \label{fig:arch}
\end{figure}

\scmacro{}'s architecture is shown in Figure~\ref{fig:arch}. The system is composed of multiple DCs; each DC contains disaggregated storage nodes and execution nodes. We consider a DC/executor to be a \textit{replica}, rather than the storage nodes. Clients can submit transactions to any DC\@. We consider transactions to be stored procedures, and the replicated content consists of parameters.

\scmacro{} is built upon our previous work \olmacro{}~\cite{maominerva2026}, an epoch-based optimistic concurrency control protocol with a highly efficient asynchronous fault-tolerant replication protocol. The main difference between \scmacro{} and \olmacro{} is that \olmacro{} uses optimistic concurrency control as the primary ordering mechanism, whereas only aborted transactions use a deterministic re-execution path. In contrast, \scmacro{}'s deterministic ordering takes priority, and many of the optimizations aim to generate an order with fewer conflicts.

The workflow of \scmacro{} roughly follows the process of the DCC protocol described in Section~\ref{sec:bg:dcc}: when a transaction is submitted to a DC, it is placed into the current global batch using a \emph{leaderless, asynchronous} replication protocol as in \olmacro{}. The batch is sealed at fixed time intervals and replicated to all DCs. Once the consensus commits the batch, replicas execute transactions in the batch in the same deterministic order. A transaction is committed and its result is returned to the client as soon as it finishes execution on one of the replicas. By the guarantees of a fault-tolerant consensus protocol, if a batch is committed on one replica, it is guaranteed to be committed on all replicas, and the execution order is also guaranteed to be the same across replicas.

\begin{figure}[h!]
    \centering
    \includegraphics[width=\columnwidth]{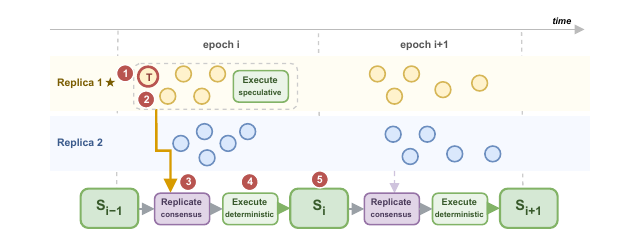}
    \caption{Workflow of \scmacro{}, viewed as a state machine over snapshots. For each transaction: \Circled{1}~submitted to its local replica, \Circled{2}~executed speculatively and batched for replication, \Circled{3}~replicated by consensus, \Circled{4}~executed in deterministic order, and \Circled{5}~returned to the client once the transaction is committed on a replica.}
    \label{fig:workflow}
\end{figure}

To achieve active-active replication, \scmacro{} synchronizes replicas by aligning the batch and epoch boundaries with the deterministic execution order. The execution of each batch creates a snapshot of the database that is consistent across replicas, and the transactions in the next batch are always executed against the snapshot of the previous batch. Deterministic sequencing of transactions is achieved by two mechanisms. The first is the logical ordering across batches, which is provided by the consensus protocol used for replication, since consensus ensures a total order. The second is the internal ordering of transactions within a batch, for which we use a locally generated sequence number at each replica together with a predefined replica priority to ensure coordination-free ordering. Figure~\ref{fig:workflow} illustrates the workflow.

Within the epoch boundaries, there is a large optimization space. Inside a deterministic batch, we no longer need to account for the interleaving of transactions across batches, so we can reorder transactions freely to reduce conflicts during the execution, since no coordination is needed as long as the process remains deterministic. We next describe the key design features.

\subsection{Speculative Execution}
\label{sec:speculation}
We first discuss the optimization during the batching phase, where transactions are executed speculatively against the latest snapshot as they arrive at the local replica. Building on the OCC execution of \olmacro{}, we introduce a Silo-based protocol~\cite{tuSilo2013}, which provides a local sequencing mechanism that serves as the deterministic order in the later execution phase rather than \olmacro{}'s local sequence numbers. The result (the write set) of each speculative execution is stored in a per-epoch buffer and treated as tentative.

\begin{figure}[h]
    \centerline{\includegraphics[width=\columnwidth] {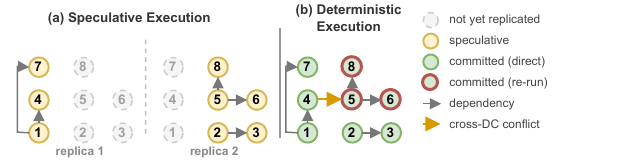}}
    \caption{(a) Before replication, replicas execute transactions speculatively against the latest snapshot. (b) Dependencies during deterministic validation: $T_4$ (replica~1) and $T_5$ (replica~2) conflict; since replica~1 has priority, $T_4$ commits directly, while $T_5$ and its dependent successors, $T_6$ and $T_8$, are discarded and rerun.}
    \label{fig:speculation}
\end{figure}

Speculative execution must be validated at the end of the batch against cross-replica read and write conflicts to ensure that the results are consistent with the deterministic execution order. This is done by building a \textit{conflict graph} during the execution phase, where each speculatively executed transaction is a vertex and the edges represent read-write or write-write conflicts that cannot be serialized. When a conflict is detected, the buffered results of the affected transaction and all of its successors are discarded. During the final execution phase, the validated results are then used directly, which skips re-execution of those transactions. An example conflict graph is shown in Figure~\ref{fig:speculation}.

Speculative execution reduces the unavoidable final execution time before commit and improves the utilization of the execution queue when contention is low. This is effective only when compute is not the bottleneck, since the added re-execution overhead may otherwise offset the gain.

\subsection{Variable Isolation Levels}
During the execution phase, \scmacro{} supports four isolation levels~\cite{adyaGeneralizedIsolationLevel2000}: \textit{Serializable} (SER), \textit{Snapshot Isolation} (SI), \textit{Read Committed} (RC), and \textit{Read Uncommitted} (RU). Each transaction declares its desired isolation level at submission time. A weaker isolation level has two benefits. First, it reduces the depth of the dependency chain, which improves throughput by increasing concurrency. Second, it allows more speculative results to remain valid, because the conflict graph contains fewer conflict edges.

We achieve this by \emph{relaxing read dependencies} during both the cross-replica conflict edge construction and the deterministic dependency chain construction, and by \textit{reordering} transactions. For each pair of transactions in the batch, \scmacro{} decides whether to add a conflict edge based on the \textit{stronger} of the two isolation levels: a write-write edge is added only if at least one transaction requires SI or stronger, and a read-write edge is added only if the \textit{reading} transaction requires SER. SI transactions are reordered to read from the snapshot rather than the write buffer, which effectively gives them an earlier order. For a write conflict between two SI transactions, the standard first-committer-wins rule is applied, which orders the transaction with the smaller sequence number first.

Because \scmacro{} only truly commits at epoch boundaries, RC transactions also read from the same snapshot; however, write-write conflicts are resolved by last-write-wins rather than by re-running the entire transaction, since RC allows the lost-update anomaly. Finally, RU transactions may read any value in the write buffer or the snapshot, and write conflicts between two RU transactions are likewise resolved by last-write-wins to establish a consistent final state. Algorithm~\ref{code:sio} summarizes the rules for dependency edge construction with variable isolation levels in \scmacro{}.

\begin{algorithm}[ht!]
    \fontsize{9pt}{10pt}\selectfont
    \caption{Dependency Edge Construction with Variable Isolation Levels}\label{code:sio}
    \begin{algorithmic}[1]
        \State Isolation strength: $\textsc{Ser} > \textsc{Si} > \textsc{Rc} > \textsc{Ru}$
        \State Each $T$ carries an isolation level $T.iso$, read-set $T.R$, write-set $T.W$, and sequence number $T.seq$
        \Function{BuildDependencyEdges}{$batch$}
            \State $E \gets \emptyset$ \Comment{conflict edges of the dependency graph}
            \ForAll{ordered pairs $(T_i, T_j)$ in $batch$ with $T_i.seq < T_j.seq$}
                \If{$T_i.W \cap T_j.W \neq \emptyset$} \Comment{write-write conflict}
                    \If{$T_i.iso \geq \textsc{Si}$ \textbf{ or } $T_j.iso \geq \textsc{Si}$}
                        \State $E \gets E \cup \{\, T_i \xrightarrow{ww} T_j \,\}$ \Comment{first-committer-wins, ordered by $seq$}
                    \Else
                        \State resolve by last-write-wins \Comment{RC/RU tolerate lost updates; no edge}
                    \EndIf
                \EndIf
                \If{$T_j.R \cap T_i.W \neq \emptyset$} \Comment{$T_j$ may read a key written by $T_i$}
                    \If{$T_j.iso = \textsc{Ser}$}
                        \State $E \gets E \cup \{\, T_i \xrightarrow{wr} T_j \,\}$ \Comment{only SER reads the write buffer}
                    \Else
                        \State $T_j$ reads from the snapshot \Comment{SI/RC/RU: ordered at batch start, no edge}
                    \EndIf
                \EndIf
            \EndFor
            \State \Return $E$
        \EndFunction
    \end{algorithmic}
\end{algorithm}

\subsection{Transaction Reordering}
\label{sec:reordering}

In addition to the SI transactions that are reordered naturally, we can also reorder some SER transactions. We apply a second type of reordering based on the optimal validation scheme of \olmacro{}.

In \olmacro{}, because of the epoch-based commitment, the validation process of its OCC execution is also performed on a batch of transactions. To reduce the number of transactions that need to be aborted, it constructs a \emph{conflict graph} to track the read-write and write-write conflicts among transactions and then applies a maximum-weight independent set (MWIS) search on the graph to find the minimal number of conflicting transactions to abort.

We adopt this mechanism and combine dependency-order determination with moving the optimal non-conflicting transaction set to the beginning of the batch, so that their speculative execution results can be used directly. This means that we \emph{reorder} the transactions that do not have conflicts in the speculative execution so that they run earlier, maximizing the number of successful speculative transactions while still maintaining a deterministic execution order.

\begin{figure}[h]
    \centerline{\includegraphics[width=\columnwidth] {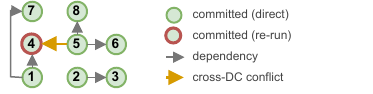}}
    \caption{Final dependency graph after reordering.}
    \label{fig:reorder}
\end{figure}

Figure~\ref{fig:reorder} shows an example using the same conflict as in Figure~\ref{fig:speculation}. Under the default replica priority, $T_4$ on replica~1 is kept while $T_5$ on replica~2 loses the conflict, which discards its entire speculative subtree $\{T_5, T_6, T_8\}$. This is wasteful, because $T_4$ is a leaf in the dependency graph, whereas $T_5$ is the root of a three-transaction subtree.

Instead of always honoring replica priority, we reorder the transactions deterministically based on the number of dependent transactions. As shown in Figure~\ref{fig:reorder}, we can keep the group of $T_5$ (weight three) over $T_4$ (weight one).

\subsection{CRDTs Integration}
\label{sec:crdt}
CRDTs are a class of specially designed data structures that support commutative operations on data items~\cite{shapiro2011crdt}. By exposing CRDTs as a primitive data type in \scmacro{}, we allow transactions to use CRDT operations to reduce write conflicts and enable fast-path commits.

When a transaction accesses a CRDT item, it applies a CRDT operation to the item rather than reading or writing the value directly. CRDT operations are designed so that they can be applied in any order and still converge to the same final state across replicas. As a result, no write-write conflict edges are created among transactions that only write to CRDT items.

When implementing a transaction with a CRDT, we replace the read--modify--write pattern with blind CRDT updates and remove the corresponding read--write and write--write dependency edges between these transactions. Our implementation currently restricts CRDT transactions to SER for simplicity, although CRDTs can be combined with weaker isolation levels by adjusting the conflict-edge construction rules in Algorithm~\ref{code:sio}.

\begin{algorithm}[ht!]
    \fontsize{9pt}{10pt}\selectfont
    \caption{Counter update transaction with and without a CRDT.}\label{code:crdt}
    \begin{algorithmic}[1]
        \Statex \textbf{Interface PN-Counter}:
        \Statex \quad $\Call{Increment}{c, n}$ \Comment{add $n$ to counter $c$}
        \Statex \quad $\Call{Decrement}{c, n}$ \Comment{subtract $n$ from counter $c$}
        \Statex \quad $\Call{Value}{c}$ \Comment{read the converged value of $c$}
        \Statex
        \Function{TxWithoutCrdt}{$key$, $\delta$} \Comment{read--modify--write}
            \State $v \gets \Call{Read}{key}$ \Comment{creates a wr-dependency on $key$}
            \State \Call{Write}{$key$, $v + \delta$} \Comment{creates a ww-conflict on $key$}
        \EndFunction
        \Statex
        \Function{TxWithCrdt}{$key$, $\delta$} \Comment{blind, commutative update}
            \If{$\delta \geq 0$}
                \State \Call{Increment}{$key$, $\delta$}
            \Else
                \State \Call{Decrement}{$key$, $-\delta$}
            \EndIf
        \EndFunction
    \end{algorithmic}
\end{algorithm}

Algorithm~\ref{code:crdt} shows an example of the difference between a transaction using a conventional counter update and a \emph{PN-Counter} (Positive--Negative Counter) update. The PN-Counter is a CRDT for integer values that supports both increments and decrements. It maintains two grow-only vectors, one for increments and the other for decrements. The value is the difference between the sums of the vectors. Each operation adds to the corresponding counter at the designated replica index, so operations issued at different replicas commute and produce the same result in any order.

\textsc{TxWithoutCrdt} reads the current value, adjusts it, and writes it back. This read--modify--write sequence creates a read--write dependency and a write--write conflict on $key$. In contrast, \textsc{TxWithCrdt} issues a blind \textsc{Increment} or \textsc{Decrement} operation, both of which are commutative. A transaction with no other conflicts can therefore use the fast-path commit described in Section~\ref{sec:eager}.

\subsection{Fast-path Commit}
\label{sec:eager}

Certain transactions can respond to the clients \textit{eagerly} as soon as they are consistently replicated, without waiting for the deterministic execution phase to complete, because it is impossible for them to be aborted. This optimization further reduces the response time for qualifying transactions. Note that fast-path commit does not change the execution semantics, since the transaction is still executed deterministically during the final execution phase; it only reduces the end-to-end response latency. We identify the following categories of transactions as eligible for fast-path commit:

\textit{\textbf{Read-only transactions}}
A read-only transaction whose reads are served entirely from the snapshot has an empty write set and can be serialized before the current batch's writers. Since the snapshot is immutable and already available locally, its speculative read results can be reused. We call this read-only optimization \emph{Snapshot Read}. It is evaluated separately from \emph{FastPath}, which enables early client acknowledgment.

\textit{\textbf{CRDT-only transactions}}
Suppose a transaction performs all of its writes as CRDT operations and serves all of its reads from the snapshot. Because CRDT operations are commutative and have no write-write conflicts by construction and snapshot reads produce deterministic results, the outcome of the transaction is fully determined at replication time. Such transactions can therefore be committed eagerly.

\textit{\textbf{RU transactions}}
A transaction running at Read Uncommitted isolation imposes no constraint on which versions it may read: any version is acceptable. Reads under RU therefore cannot cause validation failures, regardless of concurrent writes, and lost updates are permitted, so the transaction can be committed eagerly.

\begin{figure*}[t]
        \centering
        \includegraphics[width=1\textwidth]{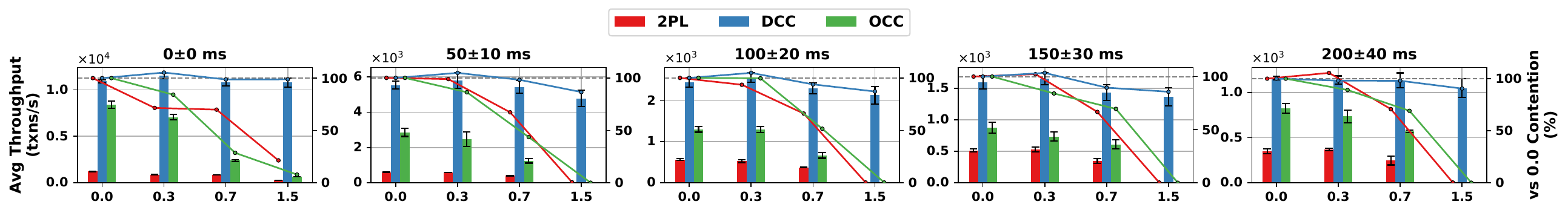}
        \caption{Saturated throughput vs.\ contention level at different cross-DC latency settings. Bars represent the absolute throughput, while lines represent the relative degradation compared to the zero-contention case.}
        \label{fig:cc-tp}
\end{figure*}

\begin{figure*}[t]
        \centering
        \includegraphics[width=1\textwidth]{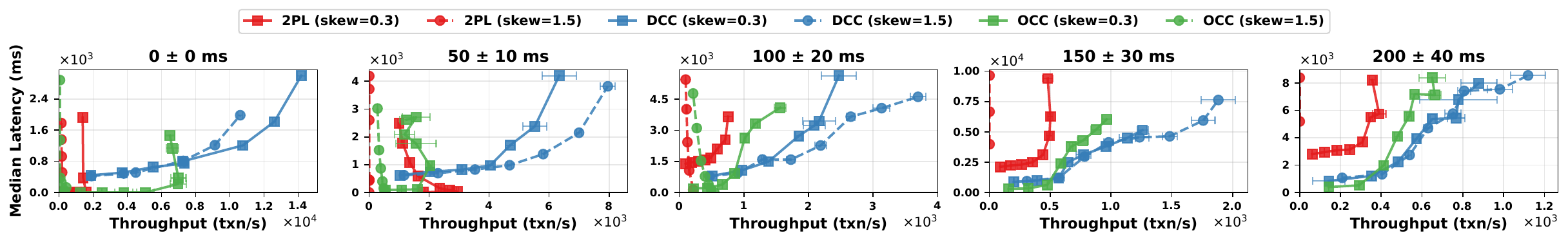}
        \caption{Throughput vs.\ median response time at different cross-DC latency settings and contention levels. OCC cannot sustain stable throughput at cross-DC latency settings of $\geq 150$\,ms and $skew=1.5$.}
        \label{fig:cc-ltvtp}
\end{figure*}

\section{Evaluation}
\label{sec:eval}
We first evaluate 2PL/OCC/DCC to test whether our model's qualitative sensitivities and bottlenecks persist in an actual implementation; then we measure the effectiveness of \scmacro{}'s optimizations and its end-to-end behavior.

\subsection{Implementation and Setup}
We build \scmacro{} on top of \olmacro{}, a transactional key-value database that implements transaction processing, replication, and benchmarks, extending it to support the additional features. The cluster is deployed on \textit{5} VMs on the cloud infrastructure of the Digital Research Alliance of Canada, where each VM is treated as a DC containing a replica of the database. Each VM has 4 vCPUs (Intel Xeon Skylake) and 48 GB of RAM and runs Ubuntu 24.04. Client threads run on a separate 16-core VM.

We implement 2PL and OCC using the same network and execution framework. We do not use consensus to implement actual distributed locking or OCC validation; instead, we use broadcast-based synchronization to simulate the coordination delay $\tau$ for a fair comparison.

\paragraph{Benchmarks and Metrics}
We measure performance using the TPC-C and YCSB benchmarks. For TPC-C, we use $100$ warehouses and a transaction mix of $50\%$ \textit{payment} and $50\%$ \textit{new-order}. For YCSB, we use the YCSB-A workload, where each transaction performs $50\%$ reads and $50\%$ blind writes with a Zipfian-distributed key-access pattern over a key space of $10^7$ keys with a $1$\,KB value size. The Zipfian factor (skew) determines the contention level, where higher values mean more key accesses are concentrated on a smaller set of hot keys, which increases contention. The transactions are implemented as stored procedures where clients only send parameters to the database.

We report the \textit{end-to-end} latency (in ms) and throughput (in txn/s), measured at the client side over \textit{all successful} transactions. The peak sustainable throughput is obtained by probing the system with an increasing number of clients until the commit throughput plateaus, which corresponds to the saturated throughput in the theoretical model. Each experiment runs for 30 seconds, and we sample client throughput every second. We report either the average throughput, after dropping the lowest 10\% of samples to remove the warm-up period, or the peak sustainable throughput. For selected experiments, we introduce artificial cross-DC latency using Linux's \texttt{tc} tool to simulate WAN conditions.

We use a Zipfian distribution instead of a uniform distribution as the models suggest to demonstrate that the model-implied sensitivities and bottlenecks persist beyond the simplifying assumptions. We therefore compare trend direction and bottleneck order, rather than calibrating the model or comparing predicted and measured point values. 

\subsection{DCC vs. OCC and 2PL}
\label{sec:cc-eval}

We compare the performance of DCC with OCC and 2PL under different cross-DC latency and contention settings using the YCSB-A benchmark. For DCC, we disable the optimizations in \scmacro{} described above for a cleaner comparison with the theoretical model. We also disable Raft consensus to match the coordination costs of 2PL and OCC.

Figure~\ref{fig:cc-tp} compares the average saturated throughput at different contention levels. Under this matched simulated-coordination setting, which isolates the paradigm effect from implementation differences, our DCC implementation in \scmacro{} outperforms 2PL and OCC in absolute throughput, confirming the caveat mentioned in Section~\ref{sec:comparison} that the protocol design itself still significantly affects actual performance. For DCC, latency still significantly affects absolute throughput because, to maintain a reasonable batch interval ($10$\,ms) for response time, the replication protocol again becomes the bottleneck, showing the necessity of optimizing the replication protocol in \scmacro{}.

Relative throughput degradation, rather than absolute throughput, is the appropriate comparison with the model, and its direction is consistent with the model-implied sensitivities. Note that, because of Zipfian skew, the skew parameter does not correspond one-to-one to the homogeneous conflict probability $p_c$ used in the model. In the experiment, increasing skew concentrates conflicts on a small set of hot keys. As a result, the measured curves are expected to drop earlier and more sharply than the smooth theoretical curves, although the relative trend remains the same. The performance of both OCC and 2PL degrades as soon as contention is set to 0.3, while DCC maintains the throughput level until 0.7. Higher latency also amplifies the degradation of 2PL and OCC, but DCC is relatively unaffected, and OCC beats 2PL when latency is higher, as expected.

Figure~\ref{fig:cc-ltvtp} illustrates the measured throughput vs.\ median response time; we gradually increase the number of clients until the systems saturate, and we report the median commit latency at the corresponding commit throughput. The measured curves are also consistent with the model's qualitative latency ordering: OCC offers better latency when load is low but degrades much more sharply than DCC as load increases, while 2PL suffers greatly in any case. When contention is higher, both OCC and 2PL also saturate almost immediately at very low input throughput, and throughput degrades as latency increases.

\subsection{Optimization Effectiveness}
\label{sec:optimization}

\begin{figure}[h]
        \centering
        \includegraphics[width=\columnwidth]{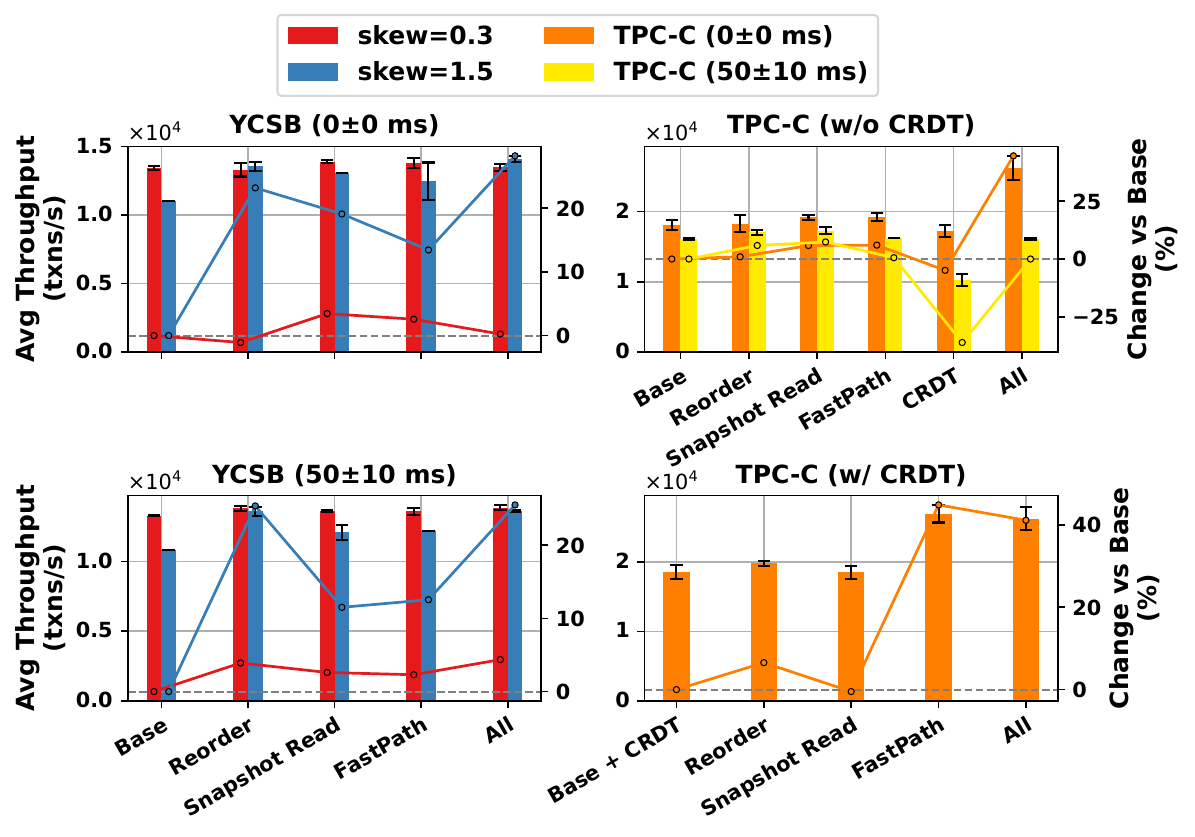}
        \caption{Ablation of optimizations at different cross-DC latency settings and contention levels. Speculative execution is enabled throughout. Bars show throughput, and lines show the percentage change in throughput relative to each panel's baseline. YCSB configurations exclude CRDT.}
        \label{fig:tp-opt}
\end{figure}

\begin{figure}[h]
        \centering
        \includegraphics[width=\columnwidth]{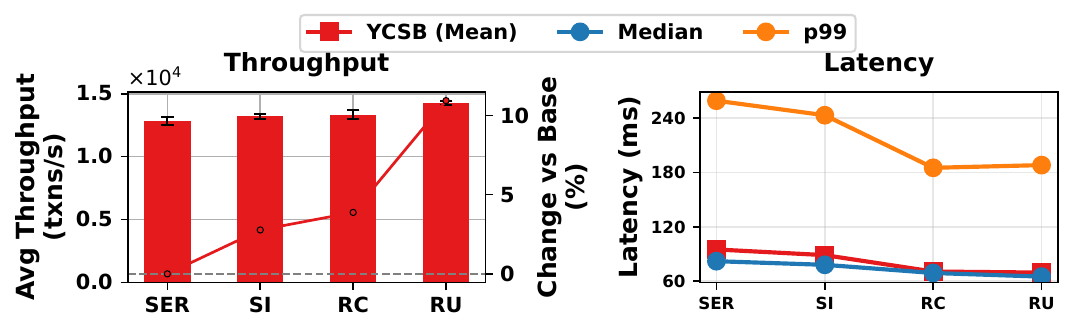}
        \caption{Saturated throughput and moderate-load latency under different isolation levels for YCSB with skew=0.3.}
        \label{fig:tp-ios}
\end{figure}

\begin{figure*}[!t]
        \centering
        \includegraphics[scale=0.5]{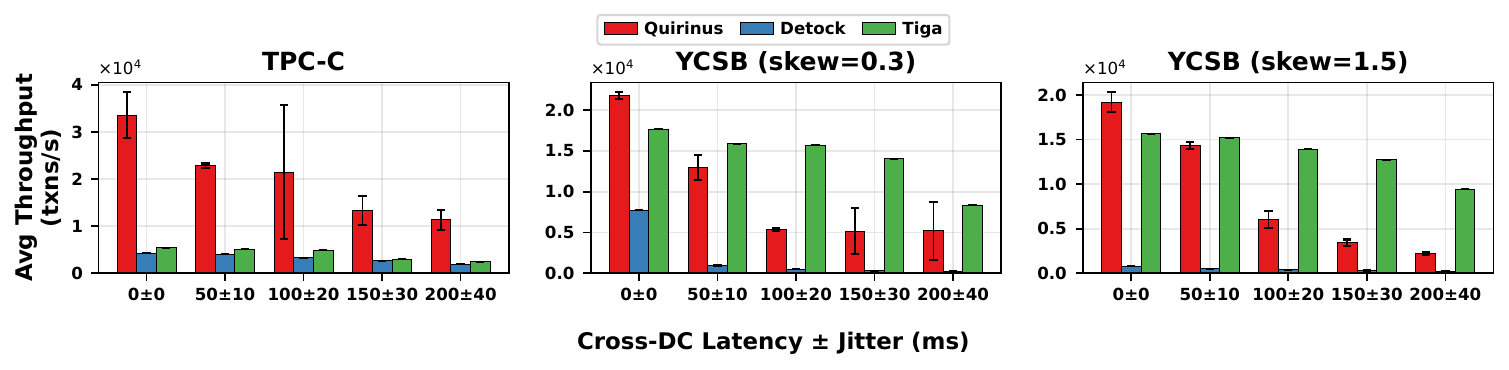}
        \caption{Peak sustainable throughput vs.\ cross-DC latency.}
        \label{fig:tp-comp}
\end{figure*}

\begin{figure*}[!t]
        \centering
        \includegraphics[scale=0.5]{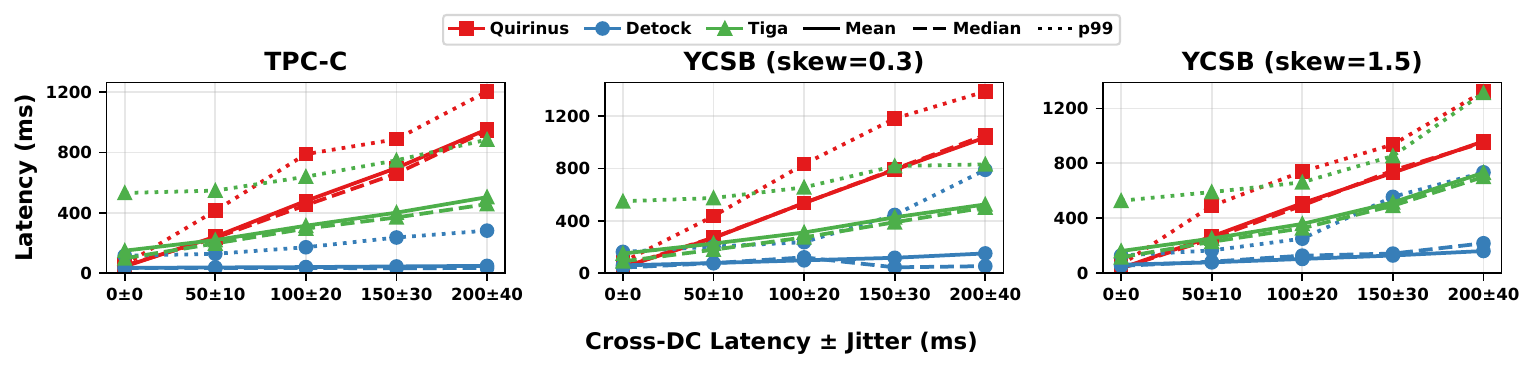}
        \caption{Transaction latency (response time) vs.\ cross-DC latency.}
        \label{fig:lt-comp}
\end{figure*}

We evaluate the throughput and latency effects of the optimizations in Section~\ref{sec:cl}. Speculative execution (Section~\ref{sec:speculation}) is enabled in every ablation configuration and defines \emph{Base}. \emph{Reorder} adds transaction reordering (Section~\ref{sec:reordering}). \emph{Snapshot Read} adds the read-only snapshot optimization, and \emph{FastPath} adds early acknowledgment (both in Section~\ref{sec:eager}). \emph{CRDT} adds CRDT integration (Section~\ref{sec:crdt}). \emph{All} combines the applicable optimizations.

The YCSB configurations exclude CRDT, including from \emph{All}. In the lower-right panel, CRDT is enabled in every configuration, and \emph{Base + CRDT} is the reference.

Figure~\ref{fig:tp-opt} illustrates the effectiveness of the individual optimization techniques. All transactions run under SER isolation, and CRDT integration is evaluated only for TPC-C. The YCSB results are shown in the two left panels. In YCSB, enabling all optimizations improves throughput by approximately $25\%$ to $30\%$ under high contention (Zipf skew of $1.5$), but by less than $10\%$ under low contention (skew of $0.3$), at both network-delay settings.

Among the individual optimizations, transaction reordering provides the largest improvement, increasing throughput by $20\%$ to $25\%$ under high contention; snapshot read-only execution and fast-path commit provide smaller gains of roughly $10\%$ to $20\%$. An interesting observation is that the gains are not additive, as the combined configuration improves throughput only slightly more than reordering alone. We hypothesize that these techniques often optimize the same transactions and overlapping portions of the critical path. The similar relative gains with and without cross-DC delay further suggest that the improvements primarily reduce contention and execution overhead rather than network coordination.

TPC-C results are shown in the upper-right panel. Unlike YCSB, the overall improvement is much more significant, but each optimization on its own is less effective, with individual techniques showing similar gains. Yet CRDT integration on its own reduces performance rather than improving it, especially in the higher-latency case ($50$\,ms and $10$\,ms jitter), where adding CRDTs alone lowers throughput by approximately $35\%$. This is likely caused by the overhead of replicating larger CRDT operations rather than only result values.

However, when combined with other optimizations, CRDTs substantially improve throughput, as shown in the bottom-right panel. In particular, combining CRDT integration with fast-path commit increases throughput by approximately $45\%$, accounting for nearly all of the gain from enabling every optimization. CRDTs alone eliminate hot write-write conflicts but leave read dependencies and execution waits, whereas the other techniques alone remain limited by the hot updates. Once CRDTs remove these update conflicts, snapshot reads remove the remaining read dependencies, reordering preserves more speculative results, and fast-path commit becomes available to update-heavy TPC-C transactions. Each technique therefore leaves a different bottleneck in place when used alone, but together they shorten the critical dependency chain enough for their benefits to outweigh their overheads.

Figure~\ref{fig:tp-ios} compares mean saturated throughput and moderate-load latency for the YCSB-A workload when all transactions are executed under the same isolation level. The experiment uses a Zipfian skew of $0.3$ and no added cross-DC delay, which isolates the effect of dependency relaxation from network coordination. For throughput, SI and RC provide small gains over SER, while RU reaches the highest throughput, roughly $10\%$ above SER. The latency results are measured at a moderate load of $500$ total clients, highlighting the benefits of weaker isolation levels. The median latency drops from SER/SI to RC/RU by about $30\%$, and the p99 latency falls more substantially, from about $230$\,ms under SER to about $180$\,ms under RC/RU.

\subsection{End-to-End Comparisons}
\label{sec:e2e}

Finally, we compare \scmacro{} with Tiga~\cite{gengTiga2025} and Detock~\cite{nguyen2023detock}, two state-of-the-art geo-distributed databases. Tiga is an OCC-based protocol that uses timestamp ordering and Paxos to agree on the commit order of transactions, and Detock uses a dependency graph and deterministic conflict resolution optimized for cross-shard operations. This experiment asks whether a fully replicated active-active database can sustain high throughput against sharded databases.

Since neither supports active-active replication, we deploy them in a sharded configuration across the 5 nodes and set the replication factor to 5 to match the full replication of \scmacro{}. For Detock, the cross-shard transactions are set to $50$\%. For \scmacro{}, we use SER isolation without optimizations enabled for a fair SER comparison. We report the \textit{peak throughput} here, which is the maximum sustainable throughput during the 30-second experiment runs.

%For Tiga, YCSB supports only a maximum of 1M keys and 5 warehouses (1 warehouse per node). 

Figure~\ref{fig:tp-comp} shows the peak throughput of the three systems and exposes the trade-off between their replication and execution strategies. On TPC-C, \scmacro{} achieves up to $6.2\times$ the peak throughput of Tiga and remains the highest-throughput design as cross-DC delay increases, although its throughput gradually declines. On YCSB, \scmacro{} leads without cross-DC delay, but Tiga degrades more gradually and becomes the highest-throughput system once delay is introduced, while Detock's throughput drops sharply. Tiga uses synchronized clocks to place most transactions on a lightweight one-RTT path~\cite{gengTiga2025}, and only shard leaders execute transactions while followers record them in their logs. This reduces coordination and replica-side computation, but does not provide active-active access. \scmacro{} remains competitive without cross-DC delay but falls behind Tiga on YCSB as delay increases, reflecting the cost of active-active multi-replica execution.

Figure~\ref{fig:lt-comp} shows that transaction latency generally increases with cross-DC delay, while Detock retains a low median for its local single-home transactions and Tiga benefits from its one-RTT fast path. In contrast, \scmacro{}'s mean and median latency track the cross-DC delay more closely, and its p99 latency reaches approximately $1.2$ to $1.4$\,s at $200$\,ms. This is the cost of \scmacro{}'s DCC design: every transaction must wait for batch formation, replication, and deterministic execution, even when it has no conflicts, and the system therefore trades per-transaction latency for throughput.

\section{Related Work}
\label{sec:related}

\textit{\textbf{Performance Modeling of OLTP Databases}}
Prior research dating back to the 1980s has developed analytical models of concurrency control, replication, and disk and network costs in transactional databases~\cite{osmanmodelsurvey2012, potier1980locking, coffmanmodel1981, nelson_analysis_1985, sheth1985analysis, thomasian19932pl, hwang_data_1996, thomasian1998cc}. However, modeling the performance trade-offs among concurrency control paradigms has received comparatively little attention, particularly under geo-replication. Recent work such as DBSeer~\cite{mozafairdbseer2013} focuses on resource prediction for specific systems rather than providing general insights. Haubenschild and Leis~\cite{haubenschild2025oltp} study the cost of cloud OLTP architectures, and GeoTP~\cite{zhuang_geotp_2025} studies lock-based geo-distributed databases.

\textit{\textbf{Geo-replicated Databases}}
Geo-distributed and replicated data\-bases have been an active research direction over the past decades~\cite{shenMakoSpeculativeDistributed2025, zhang_tapir_2018, nawabhelios2015, fanOceanVistaGossipbased2019, renslog2019, hildredCaerusLowLatencyDistributed2023, maominerva2026, gengTiga2025, zhougeogauss2023, liredblue2012, taftCockroachDBResilientGeoDistributed2020a, zhang2022starry, zhang_fast_2024, eldeeb_chardonnay_2023}. However, active-active replication remains rare. Mako~\cite{shenMakoSpeculativeDistributed2025} presents an OCC-based protocol that uses leader replication; it reduces the time transactions wait on replication by making results \textit{speculatively} visible before replication succeeds. However, its dependency tracking is local (on the leader shard) and therefore cannot address active-active replication; in addition, it relies on RDMA. GeoGauss~\cite{zhougeogauss2023} provides active-active multi-master replication, also using OCC, but it relies on a synchronous global execution barrier for each batch of transactions as an epoch. Detock~\cite{nguyen2023detock} eliminates global ordering for multi-region transactions by having participating regions independently construct the same dependency graph and resolve deadlocks deterministically; however, it retains a single home region for each item and asynchronously propagates regional logs to other regions. Starry~\cite{zhang2022starry} supports multi-master transaction processing through a semi-leader protocol that commits non-conflicting transactions at any replica on a decentralized fast path, while routing conflicts to a centralized sequencer for reordering.

\textit{\textbf{Deterministic Concurrency Control}}
After Calvin~\cite{thomsonCalvinFastDistributed2012}, one of the pioneering deterministic databases, SLOG~\cite{renslog2019} extends Calvin with locality awareness, allowing single-region transactions to commit without cross-region coordination. BOHM~\cite{faleiroBOHM2015} and Caracal~\cite{qinCaracal2021} adapt DCC to multi-version databases. Aria~\cite{luAriaFastPractical2020} uses a similar \emph{snapshot} approach, and Gria~\cite{wang_gria_2023} improves Aria with adaptive batch sizing. We also note that DCC protocols rarely view themselves as state-machine replication protocols, although they are inherently similar. Only in some blockchain systems, such as Harmony~\cite{laiHarmony2023}, which applies the same snapshot-per-block model to private blockchains, can we view determinism as part of the replication protocol. Thus, we build \scmacro{} with replication as part of the integrated design, which gives us a larger optimization space.

\section{Conclusion}
\label{sec:con}

Deterministic concurrency control is a promising approach for geo-replicated databases, but its performance under high contention and coordination delay has not been well understood. In this paper, we present a theoretical model that captures the trade-offs of DCC relative to 2PL and OCC. We also design \scmacro{}, a deterministic database optimized for active-active replication, with features such as speculative execution, transaction reordering, conflict-free data type integration, and fast-path commit. Our evaluation shows that \scmacro{} achieves high throughput on TPC-C but falls behind Tiga on YCSB as cross-DC delay increases.
\newpage
%\clearpage

\bibliographystyle{ACM-Reference-Format}
\bibliography{ref}

\appendix

\section{Derivation of Blow-up Threshold under 2PL with Geo-replication}
\label{app:blowup}

This appendix derives the lock contention threshold $P_{W,\text{MAX}}^T = 0.75/(1+\tau/a)$ in 2PL. The derivation follows Yu et al.'s high-contention 2PL approximation~\cite{yu1993cc} as an asymptotic threshold for the lock-wait fixed point.

\subsection{The Quantity Being Solved}
\label{app:yu-recap}

Yu's fixed-length 2PL model splits each lock stage into two pieces. In substate $\hat{i}$, the transaction already holds $i$ locks and executes the $i$-th access with mean time $a$. In substate $\tilde{i}$, the transaction is trying to acquire lock $i$ while still holding the previous $i{-}1$ locks.

Let $P_W$ be the probability that a lock request is blocked, and let $R_W$ be the mean lock-wait time conditional on such a block. In Yu's original model, the unconditional waiting cost per lock stage is $b = P_W R_W$. With geo-replication, every lock acquisition also pays one coordination round-trip, even if the lock is free. We therefore use
\begin{equation}
b' \;=\; \tau + P_W R_W \label{eq:bp}
\end{equation}
to represent that every request pays $\tau$, and only the contended fraction pays the extra queueing wait $R_W$.

The threshold is stated in terms of Yu's per-transaction contention
measure
\begin{equation}
P_W^T \;\equiv\; P_W(N_L-1).
\end{equation}
This is the average number of lock contentions a transaction sees over
the $N_L{-}1$ later lock requests. The first lock has no earlier locks
to hold while waiting, so Yu's blow-up calculation is naturally written
with $N_L{-}1$.

\subsection{Yu's High-contention Wait Approximation}
\label{app:limit}

Yu's blow-up threshold is obtained from the high-contention limit of
the lock-wait equation: $N_L$ is large, lock waiting dominates commit
time, and we use Yu's default residual-life correction, $f_1 = 1$, for
the remaining lock-holding time seen by a blocked requester. Under this assumption, Yu's Equation 3.19 reduces to
\begin{equation}
R_W \;\approx\; \frac{(a+b')^2 (N_L-1)}{3a}. \label{eq:Rwasy}
\end{equation}
This is the feedback step. If a transaction spends longer in each
stage, then it holds its previous locks longer; a new requester is more
likely to meet a holder with a long remaining chain of work; therefore,
the conditional wait $R_W$ grows with $(a+b')^2$.

The final replicated-commit round is still required by 2PL. It extends
the period for which all locks are held and is included in the
lock-exposure aggregate $G'$ used by the throughput calculation in the
body. It does not appear in the reduced threshold equation below because
this Yu-style high-contention limit drops commit-time constants relative
to lock waiting.

\subsection{Closing the Fixed Point}
\label{app:solve}

Equation~\eqref{eq:bp} gives another expression for the same
conditional wait:
\begin{equation}
R_W \;=\; \frac{b'-\tau}{P_W}. \label{eq:Rw-from-bp}
\end{equation}

Equating \eqref{eq:Rw-from-bp} and \eqref{eq:Rwasy} gives
\begin{equation}
\frac{b'-\tau}{P_W}
\;=\;
\frac{(a+b')^2 (N_L-1)}{3a}.
\end{equation}
Multiplying by $P_W$ and using $P_W^T = P_W(N_L-1)$:
\begin{equation}
P_W^T (a+b')^2 \;=\; 3a(b'-\tau). \label{eq:quad-pre}
\end{equation}
This is the self-consistency equation. The left side says that more
per-transaction contention makes the lock holder's remaining work more
expensive. The right side says that the only part of $b'$ that can be
explained by queueing is the amount above the unavoidable round-trip
$\tau$.

Now make the equation dimensionless:
\begin{equation}
u \;\equiv\; \frac{a+b'}{a},
\qquad
b' = a(u-1).
\end{equation}
Substitute this into \eqref{eq:quad-pre}:
\begin{equation}
P_W^T(a u)^2
\;=\;
3a\bigl(a(u-1)-\tau\bigr).
\end{equation}
After dividing by $a^2$ and rearranging, we get the quadratic
\begin{equation}
\boxed{\;
P_W^T u^2 - 3u + 3\!\left(1+\frac{\tau}{a}\right) = 0
\;}. \label{eq:quad-u}
\end{equation}
This is the quadratic used in the body of the paper. When $\tau=0$, it
reduces to Yu's original form $P_W^T u^2 - 3u + 3 = 0$.

\subsection{Reading the Threshold from the Discriminant}
\label{app:thr}

The quadratic \eqref{eq:quad-u} has coefficients
\begin{equation}
A = P_W^T,
\qquad
B = -3,
\qquad
C = 3\!\left(1+\frac{\tau}{a}\right).
\end{equation}
A finite real fixed point requires real roots, so the discriminant
must be non-negative:
\begin{equation}
B^2 - 4AC
\;=\;
9 - 12P_W^T\!\left(1+\frac{\tau}{a}\right)
\;\ge\; 0.
\end{equation}
Solving for $P_W^T$ gives
\begin{equation}
P_W^T
\;\le\;
\frac{9}{12(1+\tau/a)}
\;=\;
\frac{0.75}{1+\tau/a}.
\end{equation}
Thus the blow-up boundary is
\begin{equation}
\boxed{\;
P_{W,\text{MAX}}^T
\;=\;
\frac{0.75}{1+\tau/a}
\;}.
\end{equation}

Solving \eqref{eq:quad-u} gives
\begin{equation}
u
\;=\;
\frac{3 \pm \sqrt{9 - 12P_W^T(1+\tau/a)}}{2P_W^T}.
\end{equation}
At the blow-up boundary, the discriminant is zero and the two roots
coincide. Substituting
$P_{W,\text{MAX}}^T = 0.75/(1+\tau/a)$ gives
\begin{equation}
u_{\text{Cont}}
\;=\;
\frac{3}{2P_{W,\text{MAX}}^T}
\;=\;
2\!\left(1+\frac{\tau}{a}\right).
\end{equation}
Using $b' = a(u-1)$, the corresponding per-lock acquisition stage cost is
\begin{equation}
b'_{\text{Cont}}
\;=\;
a(u_{\text{Cont}}-1)
\;=\;
a + 2\tau.
\end{equation}
Below this threshold, the quadratic has two roots. The minus branch is the relevant low-load solution because it continuously approaches the no-contention case: as $P_W^T \to 0$, it gives $b' \to \tau$. This means the queueing wait disappears, while each lock request still pays the coordination round-trip. The plus branch has no meaningful solution in this limit.

\paragraph{Intuition}
The coordination delay does two things at once. First, it increases the per-lock stage cost from $b$ to $b'=\tau+P_W R_W$. Second, that higher stage cost increases Yu's wait equation, because lock holders keep their earlier locks longer. This feedback appears as the larger constant term $3(1+\tau/a)$ in \eqref{eq:quad-u}. A larger constant makes the discriminant disappear at a smaller value of $P_W^T$, so the system tolerates less contention before the lock-wait fixed point is unstable.

% Two checks match the intended behavior:
% \begin{itemize}
% \item If $\tau=0$, then $P_{W,\text{MAX}}^T = 0.75$, exactly Yu's
% threshold.
% \item If $\tau=a$, then $P_{W,\text{MAX}}^T = 0.375$: a round-trip as
% large as one execution stage halves the tolerable contention budget.
% \end{itemize}

\subsection{Lock Exposure at the Lock-Wait Blow-up Boundary}
\label{app:lock-exposure}

$G$ is the total time for which a transaction holds locks in Yu et al.'s paper, summed across all of its locks. With geo-replication, we use $G'$ because each serialized lock-acquisition stage pays $b'$. The transaction holds an increasing number of locks as it executes, holds its earlier locks while acquiring each later lock, and holds all $N_L$ locks during commit. Therefore,
\begin{equation}
\begin{aligned}
G'
&\;=\; \sum_{i=1}^{N_L} i\,a
      + \sum_{i=1}^{N_L-1} i\,b'
      + N_L T_{\text{Commit}} \\
&\;=\; \frac{N_L(N_L+1)}{2}\,a
      + \frac{N_L(N_L-1)}{2}\,b'
      + N_L T_{\text{Commit}}.
\end{aligned}
\label{eq:Gprime-general}
\end{equation}

At the blow-up boundary, $b'_{\text{Cont}} = a+2\tau$. Neglecting
local commit work gives
$T_{\text{Commit}} =
T_{\text{Commit}}^{\text{local}}+\tau \approx \tau$.
Substituting both values into \eqref{eq:Gprime-general}:
\begin{equation}
\begin{aligned}
G'_{\text{Cont}}
&\;=\; \frac{N_L(N_L+1)}{2}\,a
      + \frac{N_L(N_L-1)}{2}\,(a+2\tau)
      + N_L\tau \\
&\;=\; N_L^2a + N_L(N_L-1)\tau + N_L\tau \\
&\;=\; N_L^2(a+\tau).
\end{aligned}
\end{equation}
The $N_L(N_L-1)\tau$ term is the exposure caused by the
lock-acquisition rounds while earlier locks remain held. The final
$N_L\tau$ term is the additional exposure from the replicated-commit
round, during which all $N_L$ locks remain held. If local commit work
is retained, the result becomes
$G'_{\text{Cont}} = N_L^2(a+\tau)
+ N_L T_{\text{Commit}}^{\text{local}}$.

\section{Derivation of the Longest Dependency Chain Length}
\label{app:longestpath}

This section derives a closed form using the mean-field recurrence for $\ell(N)$, which is the expected length of the longest dependency chain for a newly arrived transaction $T$ given a pool of $N$ in-flight transactions ahead of it. We show that the linear ansatz $\ell(N) = AN$ is an \emph{exact} solution of the recurrence and that the slope is forced to be $A = p_c$, yielding $\ell(N) = p_c N$.

\subsection{The Recurrence}
\label{app:lp-recur}

From the model in \S\ref{sec:dcc-latency}, conditioning on $M \equiv \max\{i : T \to T_i\}$ (the highest-id out-neighbour of $T$, with $M = 0$ when $T$ has no out-edges) and applying the mean-field substitution $I_{T_i} \to \ell(i{-}1)$ gives
\begin{equation}
\ell(N) \;=\;
\sum_{m=1}^{N} p_c\,(1 - p_c)^{N - m}\,\bigl(1 + \ell(m - 1)\bigr),
\qquad \ell(0) = 0.
\end{equation}
The $m{=}0$ branch contributes $0$ to the expectation, so it does not appear in the sum. Reindexing with $k \equiv m - 1$ and writing $q \equiv 1 - p_c$ for brevity,
\begin{equation}
\ell(N) \;=\;
\sum_{k=0}^{N-1} p_c\,q^{N-1-k}\bigl(1 + \ell(k)\bigr)
\label{eq:lp-rec}.
\end{equation}

\subsection{Linear Ansatz}
\label{app:lp-ansatz}

We try a linear solution with unknown slope $A$:
\begin{equation}
\ell(N) \;=\; A\,N, \label{eq:lp-ansatz}
\end{equation}
which is consistent with the boundary condition $\ell(0) = 0$. Substituting \eqref{eq:lp-ansatz} into \eqref{eq:lp-rec} splits the right-hand side into two sums:
\begin{equation}
A\,N \;=\; \underbrace{\sum_{k=0}^{N-1} p_c\,q^{N-1-k}}_{S_1}
       \;+\; A\,\underbrace{\sum_{k=0}^{N-1} p_c\,q^{N-1-k}\,k}_{S_2}
\label{eq:lp-split}.
\end{equation}
We evaluate $S_1$ and $S_2$ in closed form below. For $S_1$, the exponents run from $N{-}1$ down to $0$, so the sum is geometric:
\begin{equation}
S_1 \;=\; p_c \sum_{k=0}^{N-1} q^{N-1-k}
       \;=\; p_c \sum_{k=0}^{N-1} q^k 
       \;=\; p_c\,\frac{1 - q^N}{1 - q} 
       \;=\; 1 - q^N \label{eq:lp-S1}
\end{equation}
using $1 - q = p_c$, which is the probability that $T$ has at least one out-neighbour in the pool of $N$.

For $S_2$, make the dependence on the upper limit explicit:
\begin{equation*}
S_2(N) \;\equiv\; \sum_{k=0}^{N-1} p_c\,q^{N-1-k}\,k.
\end{equation*}
We write $S_2(N)$ rather than just $S_2$ to compare the same weighted sum at consecutive upper limits, express the larger sum using the smaller one plus the new final term, and then verify the closed form by induction. The expression for the upper limit $N{+}1$ differs from the expression for $N$ by one new final term, $k=N$:
\begin{equation}
\begin{aligned}
S_2(N+1)
&\;=\; \sum_{k=0}^{N} p_c\,q^{N-k}\,k \\
&\;=\; \sum_{k=0}^{N-1} p_c\,q^{N-k}\,k + p_c\,N \\
&\;=\; q \sum_{k=0}^{N-1} p_c\,q^{N-1-k}\,k + p_c\,N \\
&\;=\; q\,S_2(N) \;+\; p_c\,N.
\end{aligned}
\label{eq:lp-S2-rec}
\end{equation}
The base case is $S_2(1)=0$, because the only term has $k=0$. We claim the closed form
\begin{equation}
S_2(N) \;=\; N \;-\; \frac{1 - q^N}{p_c}. \label{eq:lp-S2}
\end{equation}
Verification by induction: \eqref{eq:lp-S2} gives
$S_2(1) = 1 - (1-q)/p_c = 1 - 1 = 0$, matching the base case. For
the step, assume \eqref{eq:lp-S2} holds at $N$ and substitute into
\eqref{eq:lp-S2-rec}:
\begin{equation}
\begin{aligned}
q\,S_2(N) + p_c\,N
&\;=\; q\!\left(N - \frac{1 - q^N}{p_c}\right) + p_c N \\
&\;=\; (q + p_c)\,N \;-\; \frac{q - q^{N+1}}{p_c} \\
&\;=\; N \;-\; \frac{q - q^{N+1}}{p_c}.
\end{aligned}
\end{equation}
The target value at $N{+}1$ is
$(N+1) - (1 - q^{N+1})/p_c$. The two agree iff
$1 - (1 - q^{N+1})/p_c = - (q - q^{N+1})/p_c$, which rearranges to
$p_c = 1 - q$ --- true by definition. Hence \eqref{eq:lp-S2} holds
for all $N \ge 1$.

\subsection{Consistency Condition}
\label{app:lp-consistency}

Substituting \eqref{eq:lp-S1} and \eqref{eq:lp-S2} into
\eqref{eq:lp-split},
\begin{equation}
A\,N \;=\; (1 - q^N) \;+\; A\!\left(N - \frac{1 - q^N}{p_c}\right)
      \;=\; A\,N \;+\; (1 - q^N)\!\left(1 - \frac{A}{p_c}\right).
\end{equation}
The $A\,N$ terms cancel on both sides, leaving
\begin{equation}
(1 - q^N)\!\left(1 - \frac{A}{p_c}\right) \;=\; 0
\qquad \text{for every } N \ge 1. \label{eq:lp-consistency}
\end{equation}
For any $p_c \in (0, 1]$ the factor $1 - q^N$ is strictly positive,
so \eqref{eq:lp-consistency} forces
\begin{equation}
A \;=\; p_c. \label{eq:lp-A}
\end{equation}
The linear ansatz is therefore an \emph{exact} solution of the
mean-field recurrence \eqref{eq:lp-rec}: no $N$-dependent residual
survives, and the slope is uniquely pinned.

\paragraph{Closed form.} Combining \eqref{eq:lp-ansatz} and
\eqref{eq:lp-A},
\begin{equation}
\boxed{\;\ell(N) \;=\; p_c\,N\;}.
\end{equation}
Specialising to the in-flight pool size $N_E$ gives
$\ell(N_E) = p_c N_E$, which is the expression fed back into the
wait-time bound in the body.

\paragraph{Intuition.} Walk down the longest chain from $T$ one
hop at a time. The id-gap from a vertex of id-rank $N$ to its
highest-id out-neighbour is geometric with parameter $p_c$ on
$\{0, 1, \ldots, N-1\}$, so each hop
drops the working id by about $1/p_c$ on average. Starting from
the back of a pool of size $N$ leaves room for roughly
$N / (1/p_c) = p_c\,N$ hops before the chain hits the bottom.
\newpage

\end{document}